\documentclass[authorversion, acmtog]{acmart}

\AtBeginDocument{%
  \providecommand\BibTeX{{%
    \normalfont B\kern-0.5em{\scshape i\kern-0.25em b}\kern-0.8em\TeX}}}

\setcopyright{acmlicensed}
\acmJournal{TOG}
\acmYear{2019}
\acmVolume{0}
\acmNumber{0}
\acmArticle{0}
\acmMonth{0}
\acmDOI{https://doi.org/0000001.0000001_2.}

\usepackage{amsmath,amssymb}
\usepackage{algorithm}
\usepackage{algpseudocode}
\usepackage{bm}
\usepackage[percent]{overpic}
\usepackage{siunitx}

\DeclareMathOperator*{\argmin}{arg\,min}
\DeclareMathOperator*{\minimize}{minimize}

\DeclareMathOperator{\im}{Im}
\DeclareMathOperator{\Tr}{Tr}

\newcommand{\mca}[1]{\mathcal{#1}}
\newcommand{\mbb}[1]{\mathbb{#1}}
\newcommand{\code}[1]{\texttt{#1}}

\newtheorem{prop}{Proposition}

\renewcommand {\paragraph}[1] {\vspace{0.1in} \noindent {\textbf {#1}.}}

\begin{document}

\title{Shape Analysis via Functional Map Construction and Bases Pursuit}

\author{Omri Azencot}
\affiliation{%
  \department{Department of Mathematics}
  \institution{University of California, Los Angeles}}
\email{azencot@math.ucla.edu}
\author{Rongjie Lai}
\affiliation{%
  \department{Department of Mathematics}
  \institution{Rensselaer Polytechnic Institute}}
\email{lair@rpi.edu}

\renewcommand{\shortauthors}{Azencot and Lai}

\begin{abstract}
  We propose a method to simultaneously compute scalar basis functions with an associated functional map for a given pair of triangle meshes. Unlike previous techniques that put emphasis on smoothness with respect to the Laplace--Beltrami operator and thus favor low-frequency eigenfunctions, we aim for a spectrum that allows for better feature matching. This change of perspective introduces many degrees of freedom into the problem which we exploit to improve the accuracy of our computed correspondences. To effectively search in this high dimensional space of solutions, we incorporate into our minimization state-of-the-art regularizers. We solve the resulting highly non-linear and non-convex problem using an iterative scheme via the Alternating Direction Method of Multipliers. At each step, our optimization involves simple to solve linear or Sylvester-type equations. In practice, our method performs well in terms of convergence, and we additionally show that it is similar to a provably convergent problem. We show the advantages of our approach by extensively testing it on multiple datasets in a few applications including shape matching, consistent quadrangulation and scalar function transfer.
\end{abstract}




\begin{teaserfigure}
 \centering
 \includegraphics[width=\textwidth]{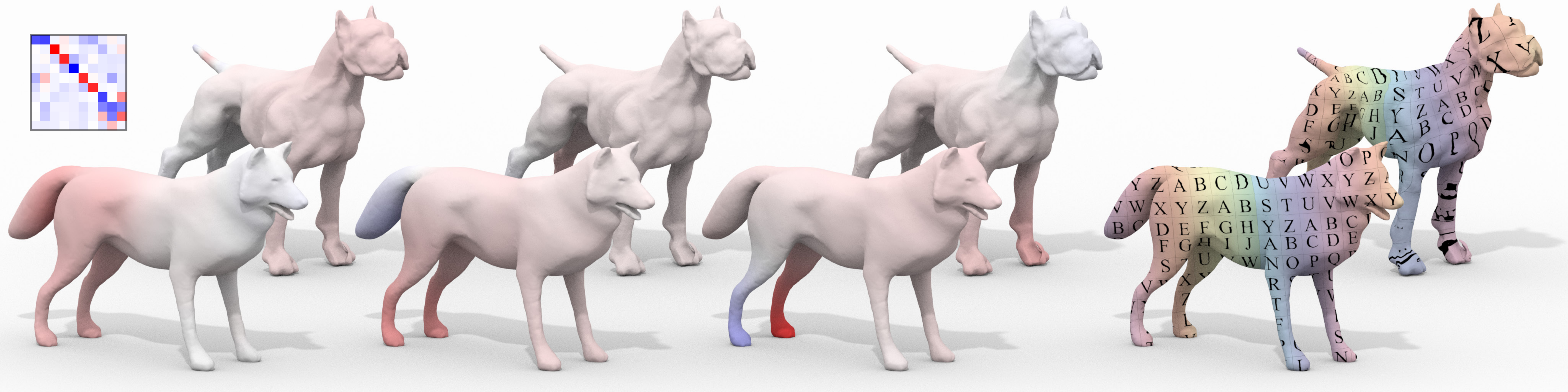}
 \caption{Given a pair of shapes and a collection of corresponding descriptors, our method produces a set of basis elements along with an associated functional map. These bases are not necessarily LB smooth and thus their aligning matrix is typically dense (left). Our machinery can be utilized in various geometry processing tasks such as non-isometric shape matching (right).}
 \label{fig:teaser}
\end{teaserfigure}

\maketitle

\newpage
\section{Introduction}

Functional maps (FM)~\cite{ovsjanikov2012functional} were recently introduced in the geometry processing community in the context of shape matching. During the last few years, FM were quickly adopted by many, serving as the key building block in a range of shape analysis frameworks. Applicative instances include mesh quadrangulation or fluid simulation tasks, in addition to the original shape correspondence problem. The goal of this paper is to propose an efficient and easy to code framework for computing improved functional map matrices.

The key idea behind FM is that instead of aligning points as in Iterative Closest Point (ICP) approaches~\cite{besl1992method}, it is often simpler to align scalar functions defined on the input shapes. Thus, a typical pipeline for computing functional maps is composed of three steps. Given a pair of shapes, one first collects a set of corresponding descriptors, such as the Wave Kernel Signature~\cite{aubry2011wave}. Second, one performs dimensionality reduction by projecting the descriptors onto a spanning subspace of basis functions. Finally, one solves an optimization problem, seeking a matrix that best aligns the projected features, possibly while minimizing additional regularizing terms.

\begin{figure*}[t]
 \centering
 \begin{overpic}[width=\textwidth]{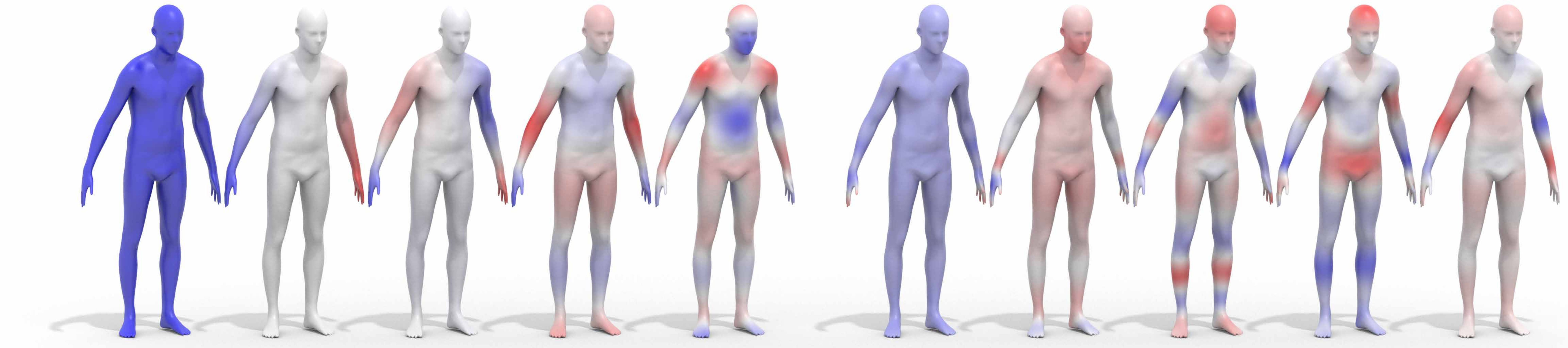}
 \put (4,21) {LB} \put (7,20) {$\mathfrak{b}_1$} \put (16,20) {$\mathfrak{b}_5$} \put (25,20) {$\mathfrak{b}_9$} \put (34,20) {$\mathfrak{b}_{11}$} \put (43,20) {$\mathfrak{b}_{26}$}
 \put (52,21) {POD} \put (56,20) {$b_1$} \put (65,20) {$b_5$} \put (74,20) {$b_9$} \put (83,20) {$b_{11}$} \put (92,20) {$b_{26}$}
 \end{overpic}
 \caption{We show above some of the modes associated with the Laplace--Beltrami (LB) operator (left) and the Proper Orthogonal Decomposition (POD) as computed for the Wave Kernel Signature~\cite{aubry2011wave} and segment information~\cite{kleiman2018robust} (right). While the LB modes encode the intrinsic geometry of the shape, they struggle with representing high frequency signals. In comparison, the POD modes faithfully represent the descriptors' spectrum while implicitly capturing the geometry as encoded in the given features.}
 \label{fig:faust_pod_vis}
\end{figure*}

Numerous extensions to the original pipeline~\cite{ovsjanikov2012functional} were proposed in the literature. These extensions can be generally classified into two research avenues. On the one hand, recent works focus on the formulation of novel regularization terms that can be incorporated into the functional map computation phase. For instance, cycle-consistency is promoted in~\cite{huang2014functional}, whereas \cite{nogneng2017informative} favor the preservation of the given descriptors. On the other hand, some papers aim to improve the functional subspaces onto which the features are being projected. For example, \cite{kovnatsky2013coupled} design basis elements to account for sign or ordering ambiguities. In this context, our work contributes in that it \emph{combines} the tasks of functional map computation and basis design into a single unified framework. Our formulation allows to harness the advancements in functional map regularization as well as to benefit from the increase in the search space of solutions when the bases are allowed to change during the optimization.

Choosing a good basis set is crucial in FM applications. In the original work~\cite{ovsjanikov2012functional}, the authors propose to utilize the spectrum of the Laplace--Beltrami (LB) operator as the spanning subspace in the second step of the pipeline. More generally, over the last few years, LB bases became the prevailing choice for function representation in many geometry processing tasks such as computing descriptors~\cite{rustamov2007laplace}, distances~\cite{solomon2014earth}, and generating shape segments~\cite{reuter2009discrete}, just to name a few. While this choice can be optimal under certain conditions~\cite{aflalo2016best}, it may sometimes lead to subpar results. To this end, Kovnatsky and others~\cite{kovnatsky2013coupled,kovnatsky2015functional,kovnatsky2016madmm,litany2017fully} optimize for joint diagonalizable (JD) basis elements to improve shape matching tasks. Solving for the bases increases the solution space, allowing to produce better correspondences.

In this paper, we address two limitations that appear in most existing work on functional maps. The first shortcoming is related to the common choice of the LB spectrum. While LB encodes the geometry of the surface, it is completely independent of the selected features, potentially introducing large representation errors. Indeed, high frequency signals such as locally supported functions will exhibit poor spectral representations~\cite{nogneng2018improved}. Thus and in contrast to previous work, our approach is based on designing basis elements that are tailored to a given collection of descriptors. In practice, we observe that employing LB-based representations often leads to the elimination of many degrees of freedom that could be re-introduced into the problem. Instead of using LB, we utilize the Proper Orthogonal Decomposition (POD) modes for dimensionality reduction purposes. POD subspaces share many of the advantageous properties of LB---they are orthonormal and have a natural ordering. However, POD modes are superior to LB in capturing high frequency data and thus improve descriptors' transfer between shapes.

The second limitation we alleviate deals with the split between the tasks of basis design and functional map computation. Indeed, most existing work focus only on one of these tasks: facilitating a fixed basis or alternatively using a closed-form solution for the functional map. Instead, we propose to merge these objectives into one larger problem. In practice, this modification leads to an increase in the search space, allowing to find better solutions for a given problem. In addition, we can independently regularize and constrain the bases and the functional map to the specific requirements of the application at hand, leading to a flexible yet effective framework.

The main contribution in this work is an effective minimization framework for computing functional basis sets on a pair of shapes and a corresponding functional map. The resulting optimization is unfortunately highly non-linear and non-convex. Nevertheless, we construct a novel and highly efficient Alternating Direction of Multipliers Method (ADMM) scheme. Specifically, our unique choice of auxiliary variables allows to naturally incorporate state-of-the-art complex regularizers promoting e.g., cycle-consistency or metric preservation. Moreover, our scheme converges empirically and we additionally show that our method is similar to a \emph{provably convergent} procedure. We evaluate our approach on several shape analysis tasks including shape matching, joint quadrangulation and function transfer. Our comparison to previous work indicates that our method achieves beyond state-of-the-art results in shape correspondences on challenging scenarios where the shapes do not share the connectivity and or only approximately isometric. 

\section{Related Work}

Functional maps~\cite{ovsjanikov2012functional} have recently gained a lot of attention in geometry processing and related fields. Some of the applications in which functional maps were found useful include shape exploration~\cite{rustamov2013map}, fluid simulation~\cite{azencot2014functional} and function transfer~\cite{nogneng2018improved}. We refer the interested reader to a recent course discussing the functional map framework and a few related applications~\cite{ovsjanikov2016computing}.

One of the main scenarios in which functional maps are employed is for computing shape correspondences between a given pair or collection of shapes. In this context, many works extend the original approach~\cite{ovsjanikov2012functional} to include various regularization terms. For instance, Nogneng and Ovsjanikov~\shortcite{nogneng2017informative} show that minimizing commutativity with descriptor operators leads to better functional maps. In~\cite{huang2014functional}, the authors promote consistency with respect to the inverse mapping, and recently, \cite{ren2018continuous} formulate orientation preserving terms into the functional maps pipeline. In addition to improving the accuracy of functional map matrices, the approaches for extracting point-to-point maps are also under development. \cite{rodola2015point} cast this problem as a probability density function estimation, whereas~\cite{ezuz2017deblurring} propose to minimize the error from projecting delta functions onto the basis and its orthogonal complement. Finally, \cite{ren2018continuous} iteratively alternate between improving the map in its spectral and spatial domains.

In all of the above works, while the functional map could be computed in any scalar basis, the eigenfunctions of the Laplace--Beltrami operator are typically used. This choice is natural given the wealth of theoretical results related to the LB spectrum, but, on the other hand, it is completely independent of the input descriptors. Recently, \cite{schonsheck2018nonisometric} proposed to design a basis via a conformal deformation while the other basis set is fixed. Probably closest to our approach is the line of work of~\cite{kovnatsky2013coupled,kovnatsky2015functional,kovnatsky2016madmm,litany2017fully} where the authors look for spectral coefficients such that the resulting basis elements are as close as possible to the LB eigenfunctions while preserving the given constraints. In contrast to their perspective, we advocate the use of a linear domain in which the descriptors are better represented, in addition to the incorporation of different regularizers. We provide a detailed comparison between our method and theirs in~\ref{subsec:cmp_to_ajd}. 

\section{Motivation and Background}

To motivate our approach, we will need the following notation. Let $\mca{M}_1 = (\mca{V}_1,\mca{F}_1)$ and $\mca{M}_2 = (\mca{V}_2,\mca{F}_2)$ be a pair of manifold triangle meshes, where $\mca{V}_1,\mca{V}_2$ are their vertex sets and $\mca{F}_1,\mca{F}_2$ are their triangle sets. We represent scalar functions using real values on vertices, i.e., $f_1:\mca{V}_1 \rightarrow \mbb{R}$ is a scalar function on $\mca{M}_1$, and similarly, $f_2:\mca{V}_2 \rightarrow \mbb{R}$ is a function on $\mca{M}_2$. Thus, $f_1$ and $f_2$ are real-valued vectors of sizes $|\mca{V}_1|=m_1$ and $|\mca{V}_2|=m_2$, respectively. We define the inner product on $\mca{M}_1$ to be
\[
\langle f_1, g_1 \rangle_{\mca{M}_1} := f_1^T G_1 g_1 \ ,
\]
where $G_1 \in \mbb{R}^{m_1 \times m_1}$ is the diagonal (lumped) mass matrix of the nodes of $\mca{M}_1$ (see e.g.,~\cite[Chap. 3]{botsch2010polygon}), and similarly, we have $\langle f_2, g_2 \rangle_{\mca{M}_2} = f_2^T G_2 g_2 $. The input to our problem is a collection of functional constraints $\{f_{1\,j}\}_{j=1}^n$ and $\{f_{2\,j}\}_{j=1}^n$ such that $f_{1\,j}$ and $f_{2\,j}$ encode the same information but on different meshes, for every $j$. Finally, we arrange the given constraints in matrices,
\[
\tilde{F}_1 = [ f_{1 1} \; f_{1 2} \; ... f_{1 n} ] \in \mbb{R}^{m_1 \times n}, \quad \tilde{F}_2 = [ f_{2 1} \; f_{2 2} \; ... f_{2 n} ] \in \mbb{R}^{m_2 \times n} \ .
\]

In its most simple form, the task of computing functional maps consists of finding a matrix $C$ that aligns the descriptors, i.e.,
\[
C \, B_1^T G_1 \, \tilde{F}_1 \approx B_2^T G_2 \, \tilde{F}_2 \ ,
\]
where $B_j$ is a basis of scalar functions on $\mca{M}_j$ for $j=1,2$. Typically, $C$ is a moderately sized $k\times k$ matrix with $k<300$. If we assume that the $G_j$ and $\tilde{F}_j$ matrices are fixed, it is natural to ask whether optimizing for $C$ \emph{and} for the $B_j$ matrices will yield improved feature matching. We show in Fig.~\ref{fig:teaser} an example of a functional map with its bases obtained in this way (left), leading to a high quality map between non-isometric shapes (right). Solving for the bases and map significantly increases the parameters from $k^2$ to $k^2\times k \cdot m_1\times k \cdot m_2$, resulting in a challenging to solve problem as $m_j$ are very large. To deal with this issue, we can consider a subspace of solutions of size $k^2\times k \cdot r_1\times k \cdot r_2$, where $r_j$ represent the spectral dimensions of some fixed bases. That is, instead of finding spatial bases, we look for spectral coefficient matrices onto predefined linear subspaces.

\begin{figure}[t]
 \centering
 \includegraphics[width=\linewidth]{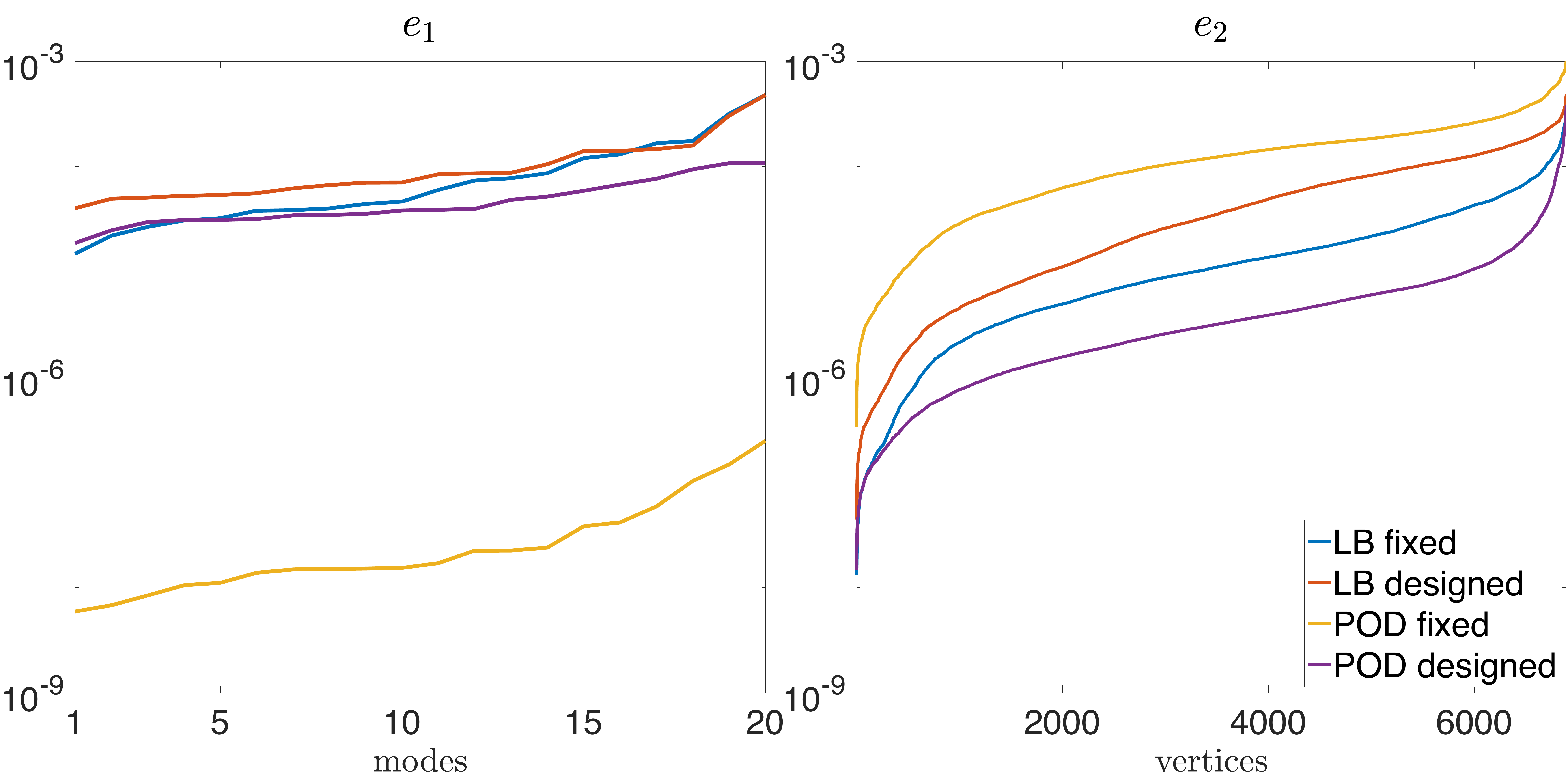}
 \caption{We plot the error distributions of feature matching when LB and POD bases are used. The above results show that designing POD modes is beneficial both in the spectral and spatial domains.}
 \label{fig:feature_err}
\end{figure}

In practice, most existing work utilize the subspace spanned by the LB eigenfunctions. In this work, we propose to change this common choice and take subspaces that better fit the given descriptors. There are several approaches in the machine learning community that could be investigated to achieve this objective. In this work, we advocate the utilization of the Proper Orthogonal Decomposition (POD) modes~\cite{berkooz1993proper}, which can be classified as a linear manifold learning method. Given a set of descriptors, the POD can be easily computed using the Singular Value Decomposition, see Sec.~\ref{sec:imp}. One of the main reasons for preferring POD modes over other bases is due to the Karhunen--Lo\'eve theorem, stating that these modes best approximate the input data under many choices of norms~\cite{xiu2010numerical}. 

Therefore, incorporating POD modes instead of the LB spectrum may be considered as a data-driven approach for representing and manipulating signals. We show in Fig.~\ref{fig:faust_pod_vis} a few modes related to the LB operator (left) and resulting from POD computation (right). One significant difference between these bases is that POD modes allow for higher frequencies when compared to a similar truncation of LB. For example, the LB $\mathfrak{b}_5$ is significantly smoother than its related POD $b_5$. Moreover, encoding descriptors in a POD subspace induces less information loss in comparison to LB representations in the context of designing bases and functional maps, as we show below.

To quantify the difference between the LB and POD subspaces, we measure the average matching error distributions per mode and per vertex, namely
\begin{align*}
e_1 = \frac{1}{n} \sum_{j=1}^n \left( C \, B_1^T G_1 \, f_{1\,j} - B_2^T G_2 \, f_{2\,j} \right)^2 \ , \\ 
e_2 = \frac{1}{n} \sum_{j=1}^n \left( B_2 \, C \, B_1^T G_1 \, f_{1\,j} - f_{2\,j} \right)^2 \ ,
\end{align*}
where the squares are taken pointwise, i.e., $e_1 \in \mbb{R}^{k}$ and $e_2 \in \mbb{R}^{m_2}$. In Fig.~\ref{fig:feature_err} we compare these errors when the bases are fixed as well as designed. Our results indicate that the fixed POD subspaces are extremely accurate for matching in the spectral domain, but yield the most error spatially. Moreover, LB bases produce poor results when designed, e.g., using joint diagonalization methods~\cite{kovnatsky2013coupled} for both error measures. Finally, designed POD modes give the most accurate estimation in the spatial domain and is second best in the spectral regime. We remark that for the POD case these errors naturally depend on the particular descriptors in use. In our applications, we employ a mixture of features such as the Wave Kernel Signature~\cite{aubry2011wave} or landmarks provided by the user.

\section{Functional Map and Basis Search (FMBS)}

Our main goal is to find basis matrices $B_1 \in \mbb{R}^{m_1 \times k}$ and $B_2 \in \mbb{R}^{m_2 \times k}$ and a functional map $C \in \mbb{R}^{k \times k}$ such that these objects best align the constraints $\tilde{F}_1$ and $\tilde{F}_2$. To reduce clutter, we scale each $\tilde{F}_j$ by its corresponding $G_j$ and denote $F_j = G_j \tilde{F_j}, j=1,2$. Formally, we consider the problem 
\begin{equation} \label{eq:fmbs} \begin{aligned}
& \minimize_{B_1,B_2,C} & & \frac{1}{2} \left| C \, B_1^T F_1 - B_2^T F_2 \right|_F^2 \\
& \text{subject to} & & B_1^T G_1 \, B_1 = I, \; B_2^T G_2 \, B_2 = I
\end{aligned} \end{equation}
where the terms $B_j^T F_j$ can be viewed as projecting the constraints onto the basis matrices. The equality conditions given by $B_j^T G_j \, B_j = I$ constrain the bases to be orthogonal with respect to the mass matrix. Unfortunately, the minimization problem~\eqref{eq:fmbs} is highly non-linear and non-convex, and thus practical solvers are challenging to construct. To alleviate these difficulties, we propose in the next section a splitting scheme that is based on the Alternating Direction Method of Multipliers (ADMM) \cite{gabay1975dual,glowinski1975approximation}.

\section{An ADMM Approach to FMBS}

The basic idea behind ADMM depends on splitting the original complex optimization task into simpler subproblems that can be solved efficiently. Under certain conditions on the objective function and constraints, it can be shown that ADMM converges. Therefore, ADMM is often the optimization framework of choice, arguably due to its computational complexity and theoretical guarantees. To allow splitting in our problem~\eqref{eq:fmbs} above, we introduce the auxiliary variables $B_1'$ and $B_2'$ and arrive at the following optimization
\begin{equation} \label{eq:fmbs_admm} \begin{aligned}
& \minimize & & \mca{E}_{\code{fid}}(B_1,B_2,C) \\
& \text{subject to} & & B_1^T G_1 \, B_1' = I, \; B_2^T G_2 \, B_2' = I, \; B_1 = B_1', \; B_2 = B_2'
\end{aligned} \end{equation}
where $\mca{E}_{\code{fid}}(B_1,B_2,C) = \frac{1}{2} \left| C \, B_1^T F_1 - B_2^T F_2 \right|_F^2$ is the \emph{data fidelity} term. We stress that while ADMM can be viewed as a standard optimization technique, the choice of auxiliary variables is highly dependent on the problem and there is no general rule for how to ``correctly'' set these variables. In particular, our unique choice leads to an empirically converging scheme for a large range of parameters, and it further allows for a natural incorporation of novel regularizers~\eqref{eq:fmbs_regs}. Finally, we mention that the auxiliary variables linearize the difficult orthogonality constraints which may lead to non-orthogonal bases in practice. However, this issue can be solved in a post-processing step. 

To minimize~\eqref{eq:fmbs_admm} we facilitate an iterative scheme $k=0,1,...$ where at each step, the unknowns are updated in an alternating style. Namely, all the variables are kept fixed except for the one which is being updated. In our case, the update order for the primal variables is $( B_1, B_2, B_1', B_2', C)$, followed by the update of the dual variables $(P_1, P_2, Q_1', Q_2')$. We note that each of the subproblems is at most quadratic in the unknown, and thus can be solved efficiently. In what follows, we discuss in detail each of the update tasks including their formulation and solution. To shorten the mathematical formulations below, we omit the step $k$ with the understanding that the variables are updated in a sequential fashion as shown in Alg.~\ref{alg:fmbs}. In addition, we denote by $\mca{L}_j(B_j,B_j',P_j,Q_j')$ the \emph{scaled} Lagrangian terms, i.e.,
\[
\mca{L}_j(B_j,B_j',P_j,Q_j') = \frac{\rho}{2} \left| B_j^T G_j B_j' - I + P_j \right|_F^2 + \frac{\rho}{2} \left| B_j - B_j' + Q_j' \right|_{\mca{M}_j}^2 \ ,
\]
where $j = 1, 2$, and $\rho \in \mbb{R}^{+}$ is a penalty parameter provided by the user and it may be updated during the optimization. 

\subsection{Updating the bases, $B_1$ and $B_2$}

The variable $B_1$ is being updated first, using the estimations of the other variables from the previous step. Specifically, we have  
\begin{align} \label{eq:B_pbm}
    B_1^{k+1} = \argmin_{B_1} \; \mca{E}_{\code{fid}}(B_1,B_2,C) + \mca{L}_1(B_1,B_1',P_1,Q_1') \ .
\end{align}
Computing the first order optimality conditions of~\eqref{eq:B_pbm} lead to a \emph{Sylvester Equation} of the form
\begin{equation} \label{eq:B1_update} \begin{gathered}
    F_1 F_1^T \, B_1 \, C^T C + \left( \rho G_1 B_1' B_1'^T G_1 + \rho G_1 \right) B_1 =  \\
    F_1 F_2^T B_2 C + \rho G_1 B_1' (I-P_1)^T + \rho G_1 (B_1'-Q_1') \ ,
\end{gathered} \end{equation}
which can be efficiently solved with numerical algorithms such as~\cite{golub1979hessenberg} implemented via e.g., \code{dlyap} in MATLAB. We emphasize that the dimensionality of Eq.~\eqref{eq:B1_update} introduces a practical challenge, as it involves dense matrices of size $m_1 \times m_1$. These concerns, along with other implementation aspects, are considered in Section~\ref{sec:imp}.

The update for $B_2$ is carried after the update of $B_1$, but before the other variables. Therefore, we use the estimate of $B_1$ at step $k+1$, whereas the rest of the variables are taken from the $k$th step. The minimization takes the following form
\begin{align} \label{eq:BT_pbm}
    B_2^{k+1} = \argmin_{B_2} \mca{E}_{\code{fid}}(B_1,B_2,C) + \mca{L}_2(B_2,B_2',P_2,Q_2') \ .
\end{align}
Problem~\eqref{eq:BT_pbm} is quadratic in $B_2$, and its solution can be computed through the following linear system
\begin{equation} \label{eq:B2_update} \begin{gathered}
    \left( F_2 F_2^T + \rho G_2 + \rho G_2 B_2' B_2'^T G_2 \right) B_2 = \\
    F_2 F_1^T B_1 C^T + \rho G_2 B_2' (I-P_2)^T + \rho G_2 (B_2'-Q_2') \ .
\end{gathered} \end{equation}

\subsection{Updating the auxiliary variables, $B_1'$ and $B_2'$}

The minimization problems associated with the unknowns $B_1'$ and $B_2'$ are similar. These optimization problems take the form
\begin{align}
    B_j'^{k+1} = \argmin_{B_j'} \frac{\rho}{2} \left| B_j^T G_j B_j' - I + P_j \right|_{\mca{M}_j}^2 + \frac{\rho}{2} \left| B_j - B_j' + Q_j' \right|_{\mca{M}_j}^2 \ ,
\end{align}
for $j=1,2$. The solution is given via the linear system
\begin{align} \label{eq:BP_update}
     \left( \rho G_j + \rho G_j B_j B_j^T G_j \right) B_j' = \rho G_j B_j (I-P_j) + \rho G_j (B_j + Q_j') \ .
\end{align}

\subsection{Updating the functional map, $C$}

Given the basis matrices $B_1$ and $B_2$, finding the best functional map that aligns the constraints in a least squares sense has a closed-form solution. Namely, we want to minimize the term $\mca{E}_{\code{fid}}(B_1,B_2,C)$ with respect to $C$, and the solution is given by
\begin{align} \label{eq:C_update}
    C^{k+1} = \left( B_2^T F_2 \right) \left( B_1^T F_1 \right)^+ \ ,
\end{align}
where $A^+$ is the pseudo-inverse of the matrix $A$.

\subsection{Updating the dual variables, $P_j$ and $Q_j'$}

The last step of our scheme is trivial and for $j=1,2,$ it is given by
\begin{equation} \label{eq:dual_update} \begin{aligned}
    P_j &= P_j + B_j^T G_j \, B_j' - I \ ,  \\
    Q_j' &= Q_j' + B_j - B_j' \ .
\end{aligned} \end{equation}

We summarize the above steps in pseudocode in Alg.~\ref{alg:fmbs}. We note that generating $\mca{O}_1^{-1}$ is computationally prohibitive as $\mca{O}$ is a \emph{large} and \emph{dense} matrix. However, we significantly reduce the computation costs by representing $B_j$ in a spectral subspace, as we discuss in Sec.~\ref{sec:imp}.

\begin{figure}[t]
  \centering
  \includegraphics[width=\linewidth]{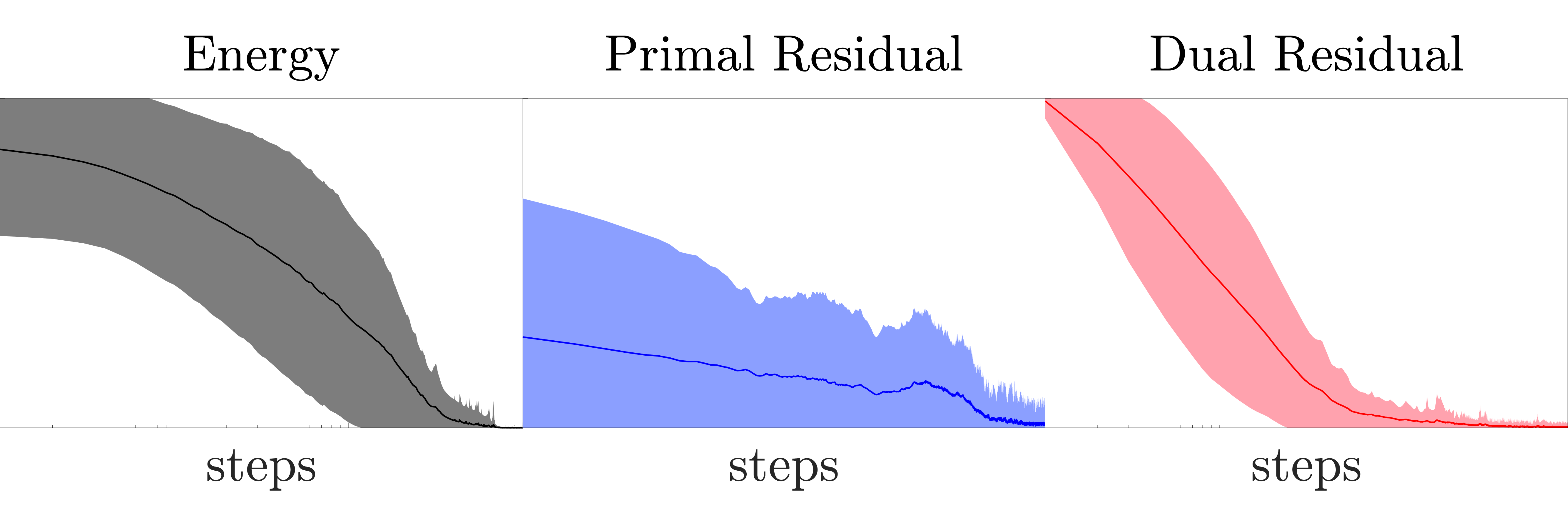}
  \caption{Our \code{fmbs} algorithm shows good empirical behavior for a large range of parameters. We compute the normalized energy and primal/dual residuals per normalized minimization step, across all our FAUST tests. The graphs above show the averaged and standard deviation of the energy (left), primal residual (middle) and dual residual (right), where all exhibit decay.}
  \label{fig:admm_stats}
\end{figure}

\begin{algorithm}
\caption{Functional Map and Basis Search (\code{fmbs})}
\label{alg:fmbs}
\begin{algorithmic}[1]

\State{Input matrices $F_1 \in \mbb{R}^{m_1 \times n} ,F_2 \in \mbb{R}^{m_2 \times n}$ and a scalar $\rho \in \mbb{R}^{+}$}
\vspace{2mm}

\State{Initialize $B_1 = B_1' = \code{SVD}(\tilde{F}_1, \, k), \, B_2 = B_2' = \code{SVD}(\tilde{F}_2, \, k)$}
\Statex{\hspace{11mm} $P_1 = 0, \, P_2 = 0, \, Q_1' = 0, \, Q_2' = 0$}
\vspace{2mm}

\For{$k=0,1,2,...$}

\State{Solve $B_1 = \code{dlyap}(-\mca{O}_1^{-1}\mca{A}_1,\mca{B}_1,\mca{O}_1^{-1}\mca{C}_1) \ ,$ where} \Comment{Eq.~\eqref{eq:B1_update}}
\Statex{\hspace{8mm} $\mca{A}_1 = F_1 F_1^T$}
\Statex{\hspace{8mm} $\mca{B}_1 = C^T C$}
\Statex{\hspace{8mm} $\mca{C}_1 = F_1 F_2^T B_2 C + \rho G_1 B_1' (I-P_1)^T + \rho G_1 (B_1'-Q_1')$}
\Statex{\hspace{8mm} $\mca{O}_1 = \rho G_1 B_1' B_1'^T G_1 + \rho G_1$}
\vspace{2mm}

\State{Solve $\mca{A}_2 B_2 = \mca{B}_2 \ ,$ where} \Comment{Eq.\eqref{eq:B2_update}}
\Statex{\hspace{8mm} $\mca{A}_2 = F_2 F_2^T + \rho G_2 + \rho G_2 B_2' B_2'^T G_2$}
\Statex{\hspace{8mm} $\mca{B}_2 =  F_2 F_1^T B_1 C^T + \rho G_2 B_2' (I-P_2)^T + \rho G_2 (B_2'-Q_2')$ }
\vspace{2mm}

\State{Update $B_1'$ by solving Eq.~\eqref{eq:BP_update} with $j=1$}
\vspace{2mm}

\State{Update $B_2'$ by solving Eq.~\eqref{eq:BP_update} with $j=2$}
\vspace{2mm}

\State{Solve $C = \left( B_2^T F_2 \right) \left( B_1^T F_1 \right)^+$} \Comment{Eq.~\eqref{eq:C_update}}
\vspace{1mm}

\State{Update the dual variables using Eq.~\eqref{eq:dual_update}}
\vspace{1mm}

\State{Update $\rho$ following Section 3.4 in~\cite{boyd2011distributed} }
\EndFor

\end{algorithmic}
\end{algorithm}

\subsection{Provably convergent FMBS} 

Most convergence results related to ADMM handle problems with convex objective functions and linear constraints. Recently, \cite{WangYinZeng2015_global} and \cite{gao2018admm} extended the convergence analysis of ADMM to a significantly larger class of problems including non-convex objective terms and non-linear constraints. In particular, in the latter work, the authors investigate the case where \emph{biaffine} constraints are given, namely, constraints involving two variables which become linear when one variable is kept fixed. For instance, our orthogonality conditions $B_j^T G_j B_j' = I$ are exactly of this form. Moreover, \cite{gao2018admm} relax the convexity requirements on the objective function and allow to include differentiable terms instead. 


In practice, Alg.~\ref{alg:fmbs} behaves well and it exhibits energy decrease for many choices of parameters as we show in Fig.~\ref{fig:admm_stats} and in Sec.~\ref{sec:eval}, however, it does not satisfy the conditions given in~\cite{gao2018admm}. To show convergence, we consider in App.~\ref{app:prop_proof} a different minimization~\eqref{eq:fmbs_admm2} for which we can show the following result.
\begin{prop} 
Under some mild conditions, problem~\eqref{eq:fmbs_admm2} satisfies all the requirements in~\cite{gao2018admm} and thus its ADMM converges to a constrained stationary point. That is, the sequence of variable updates $\left\{ \mca{X}^k, \mca{Z}^k \right\}_{k=0}^\infty$ is bounded and every limit point $(\mca{X}^*,\mca{Z}^*)$ is a constrained stationary point.
\end{prop}

\begin{figure*}[t]
  \centering
  \includegraphics[width=\textwidth]{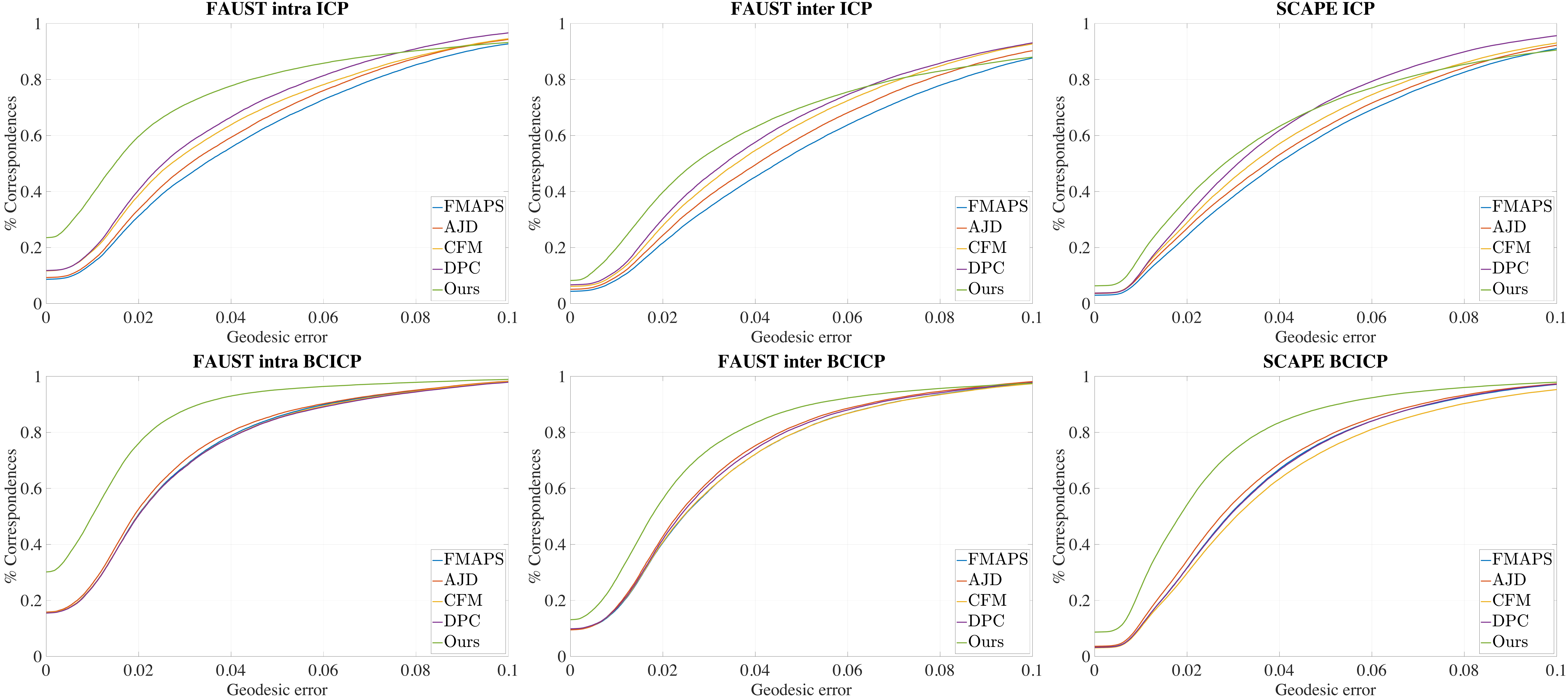}
  \caption{We compute the geodesic error between the ground-truth mapped vertices to those mapped with the above methods, and we accumulate the amount of points that reached a certain error~\cite{kim2011blended} and visualize it in the above graphs. Noticeably, our method yields the best averaged error over other methods for both ICP and BCICP map extraction. We note that basis design methods such as ours and AJD~\cite{kovnatsky2013coupled} gain the most from advanced mapping methods such as BCICP~\cite{ren2018continuous}.}
  \label{fig:lmk_wks_cull_err}
\end{figure*}

\subsection{Regularized FMBS}

One of the key aspects of our minimization~\eqref{eq:fmbs_admm} is that it introduces many degrees of freedom via the unknowns $B_1, B_2$ and $C$. While in general it is a positive feature of our approach, the associated optimization requires a significant amount of descriptors $n$. To relax this dependency, we propose to incorporate regularization terms into our problem. In particular, we add a consistency regularizer that takes into account the inverse functional map $D$. Moreover, we add an isometry promoting term which is given by commutativity with the LB operator \cite{ovsjanikov2012functional} and Dirichlet energies that favor smooth basis elements. We note that other regularizers such as descriptor commutativity~\cite{nogneng2017informative} or orientation preservation~\cite{ren2018continuous} may be also considered. Formally, we propose the following objective function
\begin{equation} \begin{aligned} \label{eq:fmbs_regs}
    & \mca{E} = \mca{E}_{\code{fid}} + \mu_{\code{cfid}} \mca{E}_{\code{cfid}} + \mu_{\code{iso}} \mca{E}_{\code{iso}} + \mu_{\code{dir}}\mca{E}_{\code{dir}} \ , \\
    & \mca{E}_{\code{cfid}} = \frac{1}{2} | B_1^T F_1 - D \, B_2^T F_2 |_F^2\ , \\
    & \mca{E}_{\code{iso}} = \frac{1}{2} | C \, B_1^T W_1 \, B_1' - B_2^T W_2 \, B_2' \, C |_F^2\ , \\
    & \mca{E}_{\code{dir}} = \frac{1}{2} \Tr \left( B_1^T W_1 B_1' \right) + \frac{1}{2} \Tr \left( B_2^T W_2 B_2' \right) \ ,
\end{aligned} \end{equation}
where $\mu_{\code{cfid}}, \mu_{\code{iso}}, \mu_{\code{dir}} \in \mbb{R}^{+}$ are penalty scalars, $\Tr$ yields the trace of a matrix, and $W_j$ is the cotangent weights matrix~\cite{pinkall1993computing} of shape $\mca{M}_j$ for $j=1,2$. One of the key aspects of our framework resulting from our ADMM formulation~\eqref{eq:fmbs_admm} is that it allows to combine challenging regularizers~\eqref{eq:fmbs_regs} in a straightforward way. Thus, the formulation of the minimization that uses $\mca{E}$ and its associated ADMM is somewhat technical, and we defer the derivation to the supplementary material.

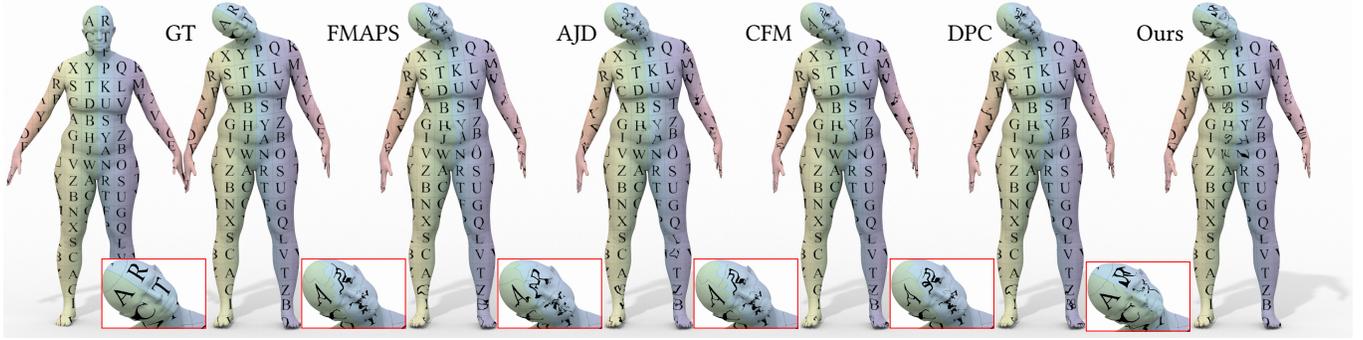
\begin{figure*}[t]
  \centering
  \begin{overpic}[width=\textwidth]{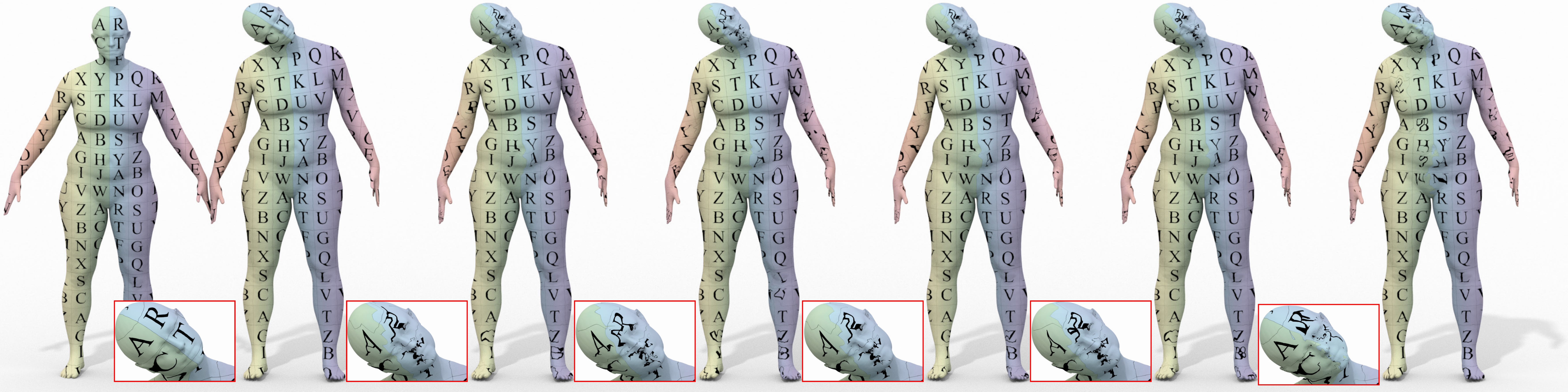}
  \put (12,22) {GT} \put (24,22) {FMAPS} \put (41,22) {AJD} \put (55,22) {CFM} \put (70,22) {DPC} \put (84,22) {Ours}
  \end{overpic}
  \caption{We visualize the quality of the computed correspondences using texture transfer. We note that the alternative methods struggle with elongated or small areas as the head and legs, whereas our method achieves improved results as can be compared to the ground-truth (GT).}
  \label{fig:cmp_map_tex}
\end{figure*}

\section{Implementation Details}
\label{sec:imp}

In what follows we describe a few technical aspects related to our method including dimensionality reduction of problem~\eqref{eq:fmbs_admm}, variable initialization, stopping condition and the development platform.

\paragraph{Dimensionality reduction} Solving~\eqref{eq:fmbs_admm} directly is computationally prohibitive when the shapes consist many vertices. To overcome this difficulty, we propose to reduce the spatial dimension and use a spectral domain instead, allowing for fast computation times while retaining a significant amount of degrees of freedom. Specifically, we take the left singular vectors obtained by computing the Singular Value Decomposition (SVD) of the given constraints. Namely,
\[
\tilde{U}_j \tilde{S}_j \tilde{V}_j^T= \code{SVD} \left( \tilde{F}_j \right) \ , \quad j=1,2 \ ,
\]
where we denote $U_j = \tilde{U}_j(: \,,1:r)$ the $r$ most significant modes. In our experiments we choose $r$ such that $U_j$ covers at least $90\%$ of the spectrum. We note that other spectral bases could be considered, e.g., the LB basis itself~\cite{kovnatsky2013coupled}. However, each choice leads to a different optimization having its own set of assumptions and challenges. In Sec.~\ref{sec:eval}, we compare our approach to other methods.

To incorporate $U_j$ into our optimization, we denote the changes in boldface and perform the following modifications, 
\begin{align} \label{eq:dim_red}
\bm{B}_j = U_j^T B_j \ , \quad \bm{F}_j = U_j^T F_j \ , \quad \bm{G}_j = U_j^T G_j \, U_j \ , \quad j=1,2 \ ,
\end{align}
yielding matrices of sizes $r_j \times k$, $r_j \times n$ and $r_j \times r_j$, respectively. Substituting the above components with their high dimensional counterparts is the only change needed to obtain a spectral version of Alg.~\ref{alg:fmbs}. Finally, given $\bm{B}_j$, we reconstruct $B_j$ via $B_j = U_j \, \bm{B}_j$. We note that while our approach strongly depends on the input features for deriving the low-dimensional subspaces, in our experiments we observed that it works quite well with a variety of descriptors such as WKS and segmentation information.


\paragraph{Variable initialization and stopping rule} In our tests we noticed that our method is robust to the choice of initial values. Nevertheless, we describe the particular values we used in our experiments. To initialize the primal variables $B_j$ and $B_j'$, we take the first $k$ singular vectors of the respective descriptors, $\tilde{F}_j$. This computation is denoted by $\code{SVD}(\tilde{F}_j,k)$ in Alg.~\ref{alg:fmbs}. Using these bases, we can solve Eq.~\eqref{eq:C_update} to obtain an initial $C$. The dual variables $P_j$ and $Q_j'$ are set to zero matrices of an appropriate size. The stopping condition we used is based on the primal and dual residuals and is detailed in~\cite[Sec. 3.3]{boyd2011distributed}, where the maximum number of steps is $10,000$.

\begin{table}[b]
\begin{tabular}{ccccc}
 \hline
 Dataset & r & $\mu_{\code{cfid}}$ & $\mu_{\code{iso}}$  & $\mu_{\code{dir}}$ \\
 \hline
 FAUST intra & 0.9 & \num{1e-4} & \num{1e-6}  & \num{1e-2} \\
 FAUST inter & 0.9 & \num{1e-4} &  \num{1e-5}  & \num{1e-4} \\
 SCAPE & 0.9 & \num{1e-3} & \num{1e-6} & \num{1e-4}  \\
 Remeshed FAUST intra & 0.99 & \num{1e-2} & \num{1e-5}  & \num{1e-6}  \\
 Remeshed FAUST inter & 0.99 & \num{1e-2} & \num{1e-5} & \num{1e-6} \\
 Remeshed SCAPE & 0.99 & \num{1e-1} & \num{1e-5} & \num{1e-6} \\
 \hline
\end{tabular}
\caption{The parameter values used in our tests for each dataset.}
\label{tab:param}
\end{table}

\begin{figure}[b]
  \centering
  \includegraphics[width=\linewidth]{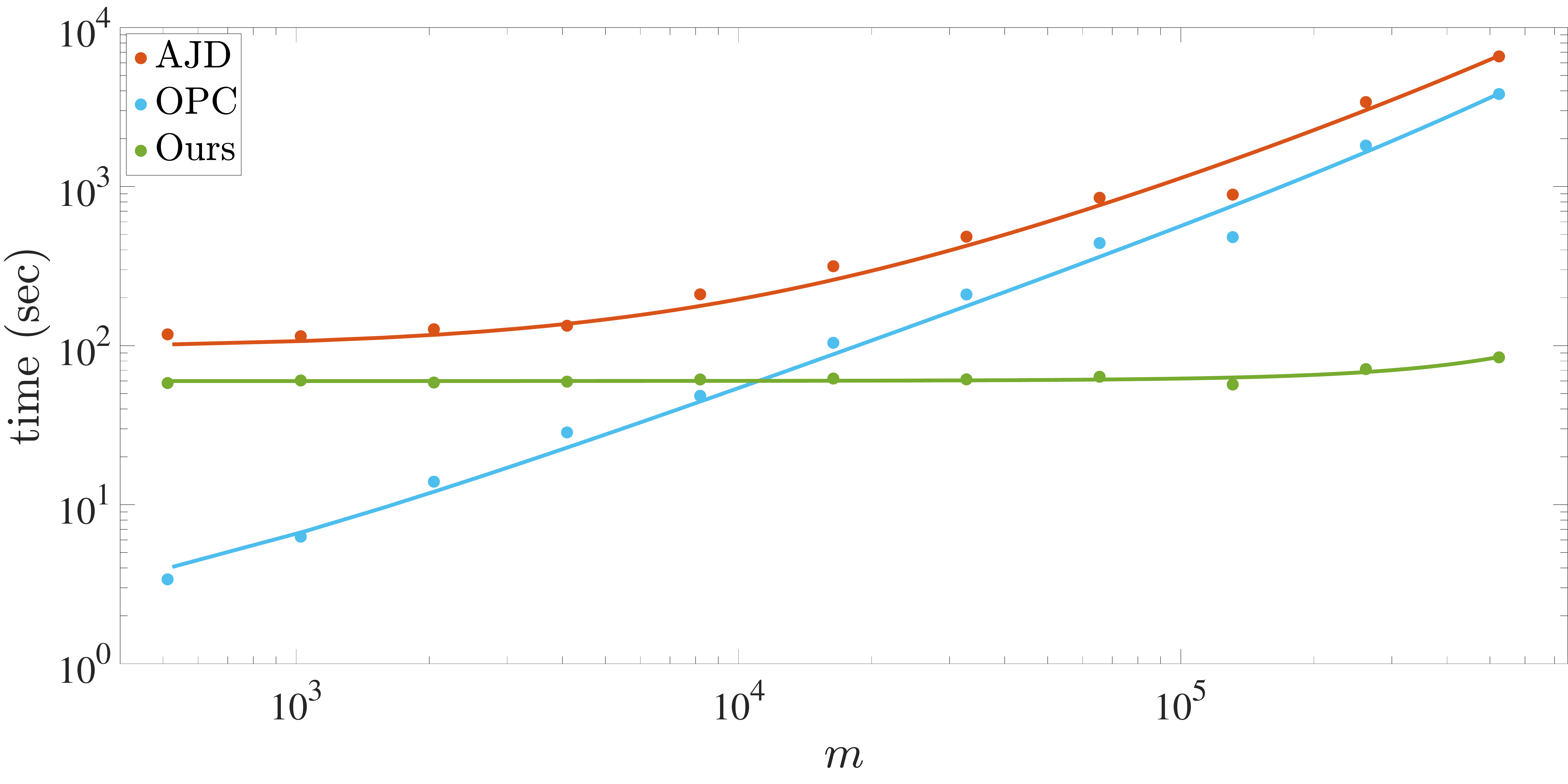}
  \caption{We compare the total pre-processing and computation times of the above methods on a pair of shapes for a large range of vertex counts, $m$. Our method is significantly faster than AJD and OPC for high vertex counts, where for low number of vertices OPC is more efficient than our approach.}
  \label{fig:timings_graph}
\end{figure}

\paragraph{Development platform and parameters} We implemented our method in MATLAB, using its built-in optimization tools such as \code{dlyap} and \code{mldivide}. Our approach was tested on an Intel Core i7 2.6GHz processor with 16GB RAM. We show in Fig.~\ref{fig:timings_graph} a runtime comparison to AJD~\cite{kovnatsky2013coupled} and OPC~\cite{ren2018continuous} on meshes of sizes $1k-500k$ vertices. The parameters of our method include the penalty scalars $\mu_\code{cfid}, \mu_\code{iso}$ and $\mu_\code{dir}$ for the different energy terms~\eqref{eq:fmbs_regs}. We list our choices in Tab.~\ref{tab:param}, which also shows how much of the spectrum we employ, given by the $r$ parameter. Finally, the size of the functional map and the associated bases was $k=20$ unless noted otherwise.

\section{Evaluation and Results}
\label{sec:eval}

To evaluate our method, we consider several applications in which functional maps are useful such as extraction of point-to-point maps~\cite{ovsjanikov2012functional}, function transfer~\cite{nogneng2018improved}, and consistent quadrangulation~\cite{azencot2017consistent}. We test our approach on a variety of datasets including SCAPE~\cite{anguelov2005correlated}, TOSCA~\cite{bronstein2008numerical} and FAUST~\cite{bogo2014faust}, and SHREC07~\cite{giorgi2007shape}. In our comparison, we consider other methods for computing functional maps such as functional maps (FMAPS)~\cite{ovsjanikov2012functional}, approximate joint diagonalization (AJD)~\cite{kovnatsky2013coupled}, coupled functional maps (CFM) \cite{eynard2016coupled}, descriptor preservation via commutativity (DPC)~\cite{nogneng2017informative} and orientation preserving correspondences (OPC)~\cite{ren2018continuous}. In our comparison, we only use the functional map matrices as computed using the above techniques, and we discard any other improvements related to a specific application. In all cases, we used the authors' recommended parameters or we searched for the best ones. 

\subsection{Extracting point-to-point maps}

One of the main applications of functional maps is the computation of point-to-point correspondences between pairs of shapes. In our comparison, we consider two different scenarios. The first includes the original FAUST and SCAPE shapes using $20$ landmarks and $100$ Wave Kernel Signature (WKS)~\cite{aubry2011wave} features. In the second case, we remesh the shapes and use consistent segmentation data~\cite{kleiman2018robust} with WKS descriptors. We emphasize that the latter scenario is extremely challenging as it is completely automatic, it involves approximate features, and the meshes have different connectivities. The pairs we use appeared previously in~\cite{kim2011blended,chen2015robust}. For map extraction we employ the ICP method proposed in~\cite{ovsjanikov2012functional} and the recent BCICP approach~\cite{ren2018continuous}, although other methods~\cite{rodola2015point,ezuz2017deblurring} could be used. Our evaluation metrics include the computation of cumulative geodesic errors~\cite{kim2011blended} and visualization of transferred scalar functions or textures.

In Fig.~\ref{fig:lmk_wks_cull_err} we show the average cumulative geodesic errors of the first scenario. We note that our approach achieves a significant improvement over all the other competing methods. In particular, when ICP extraction is facilitated, our method yields very good results on FAUST intra which involves pairs of different poses of the same people. Interestingly, our method benefits the most from recent advances in map extraction techniques~\cite{ren2018continuous} as can be seen in the second row. Specifically, using BCICP increases the gap between our results vs. others on FAUST inter (different people, different pose) and SCAPE. This hints that our functional map and associated bases introduce more degrees of freedom which could be exploited in elaborated methods such as~\cite{ren2018continuous}. This behavior can be additionally seen in AJD~\cite{kovnatsky2013coupled} BCICP results which surpass most methods even though their ICP measures were lower than others in general. We do not compare to OPC~\cite{ren2018continuous} in this setup as we use non-symmetric landmarks and thus there is no advantage in using their orientation preserving regularization over, e.g., DPC~\cite{nogneng2017informative}.

\begin{figure}[t]
  \centering
  \begin{overpic}[width=\linewidth]{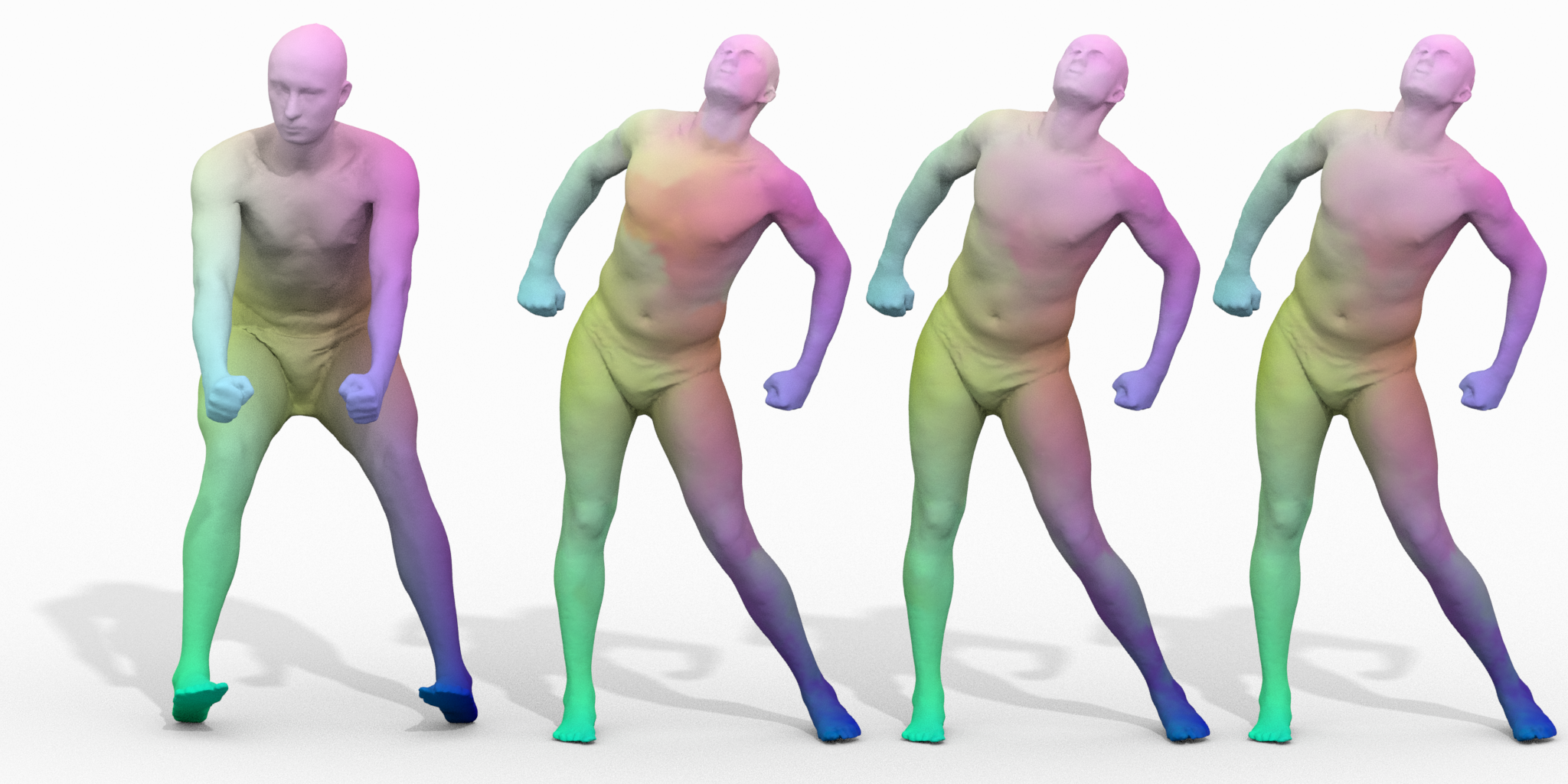}
  \put (34,45) {$\mca{E}_{\code{fid}}$} \put (55,45) {+$\mca{E}_{\code{iso}}$} \put (78,45) {+$\mca{E}_{\code{dir}}$}
  \end{overpic}
  \caption{To compare the advantages of regularization, we solve our problem with various combinations of energy terms. Using the obtained maps, we transport the coordinate functions from the target to the source and show the results above. We achieve a significant improvement when we regularize as can be seen around the chest and head (middle-right and right) vs. the non-regularized result (middle-left).}
  \label{fig:cmp_energy_terms}
\end{figure}

We demonstrate the error measures of Scenario $2$ in Fig.~\ref{fig:seg_wks_cul_err}. We stress that this setup is particularly challenging as the shapes do not share the connectivity and we use automatically computed features. Nevertheless, our method exhibits the best results on FAUST both for the isometric and non-isometric cases when ICP map extraction is applied. Moreover, when we utilize BCICP on FAUST our method and OPC yield the best scores compared to the alternative methods. Finally, the remeshed SCAPE was an extremely difficult test case, leading to mappings of poor quality in general for most methods (notice the $y$-axis gets to $0.6$ instead of $1$). For this dataset, CFM and DPC produced good measures for ICP, and our method and OPC were the highest with BCICP refinement.

The point-to-point correspondence allows to map information from the target to the source. In Fig.~\ref{fig:cmp_map_tex}, we compare the mappings generated in Scenario $1$ on a single pair of FAUST intra using texture transfer. The meshes in this dataset are in $1-1$ correspondence and thus we can use the ground-truth (GT) map for comparison. Overall, the performance of the tested methods was generally good. However, small parts of the body such as hands and legs were less accurate for FMAPS, AJD and DPC. Moreover, other methods exhibit large errors in the head, whereas ours correctly finds the symmetry line (see the zoom below). 

\subsection{Effect of regularization}

To evaluate the benefits of utilizing regularizing terms, we visualize the map quality via coordinate function transfer in Fig.~\ref{fig:cmp_energy_terms}. Indeed, there is a clear improvement when $\mca{E}_{\code{iso}}$ is introduced (middle right) vs. using $\mca{E}_{\code{fid}}$ alone (middle left) as can be seen on the chest and head. Adding $\mca{E}_{\code{dir}}$ (right) is not beneficial in this case as it is visually indistinguishable from the $\mca{E}_{\code{iso}}$ (middle right) case.

\begin{figure}[t]
  \centering
  \includegraphics[width=\linewidth]{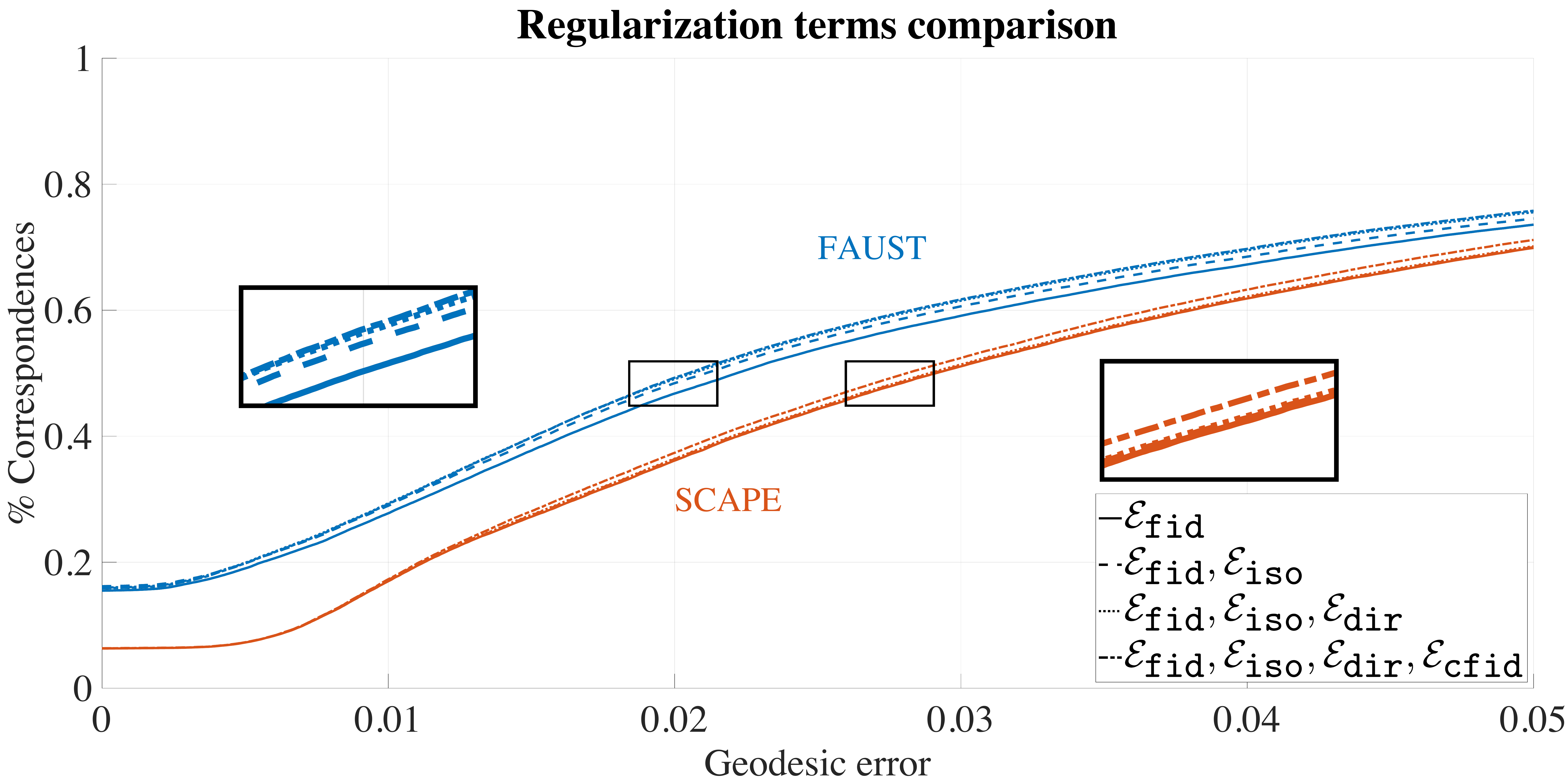}
  \caption{We plot the cumulative geodesic error for maps computed using various regularization settings. Our results indicate that the regularized problems yield better correspondences. See the text for additional details.}
  \label{fig:cmp_energy_terms_graph}
\end{figure}

In addition to this visualization, we also run our algorithm on FAUST and SCAPE in scenario 1, using different regularization configurations. We show in Fig.~\ref{fig:cmp_energy_terms_graph} the cumulative geodesic error of these tests. For each dataset the solid line represents using only $\mca{E}_{\code{fid}}$, the dashed is the result when we incorporate $\mca{E}_{\code{iso}}$, the dotted line is produced by adding $\mca{E}_{\code{dir}}$, and we get the dash-dot line by minimizing the full $\mca{E}$. On average, we observe a significant gain when regularization is used (see also the zoomed plots in Fig.~\ref{fig:cmp_energy_terms_graph}). In particular, the consistency term (dash-dot line) helps both with respect to the accuracy of the results and the empirical convergence of the problem. We note that the Dirichlet penalization (dotted line) improves the results of FAUST, whereas for SCAPE its contribution is less apparent.

\begin{figure}[t]
  \centering
  \begin{overpic}[width=\linewidth]{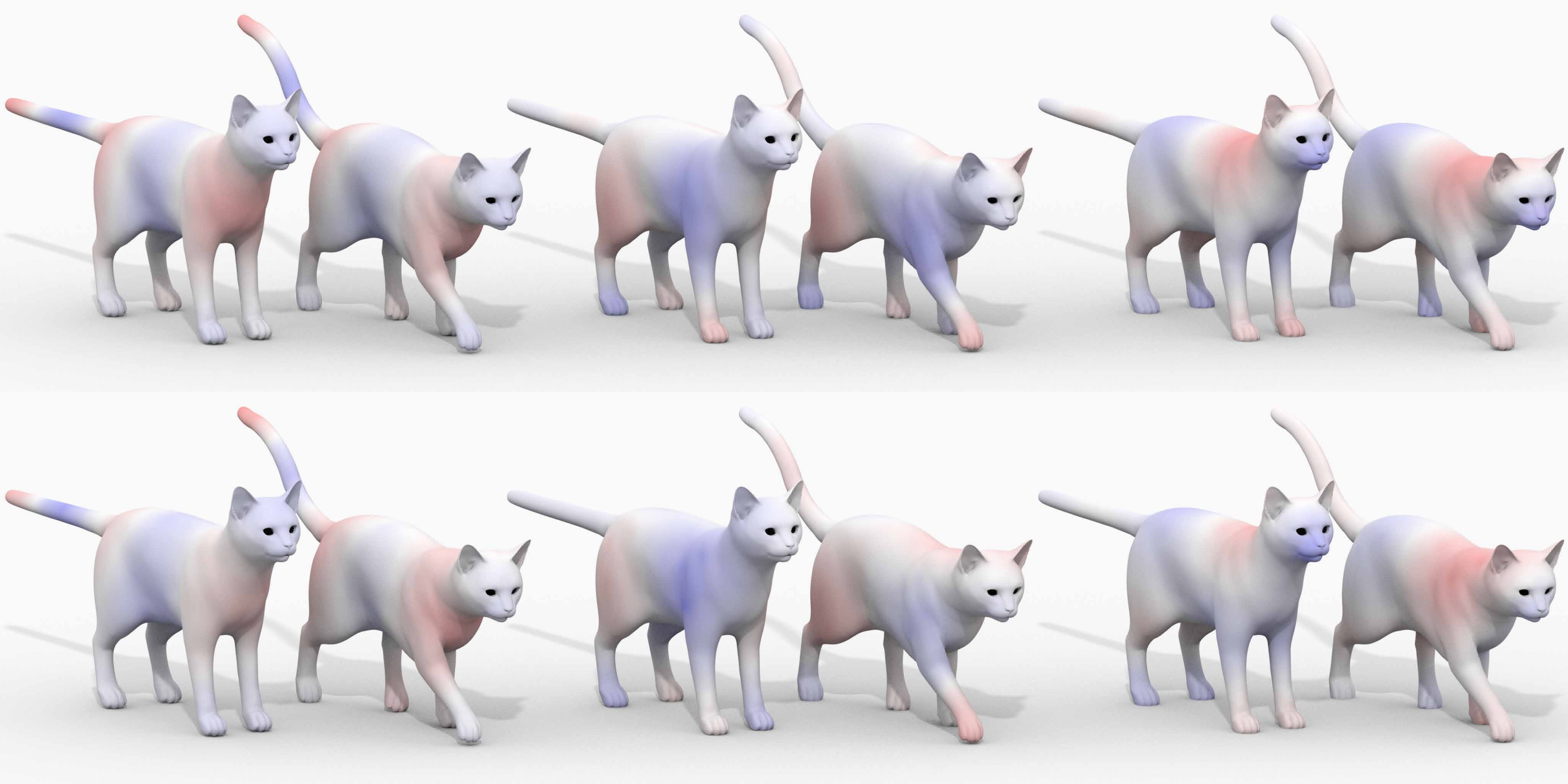}
  \put (5,45) {AJD} \put (5,20) {Ours}
  \end{overpic}
  \caption{We show above the first three basis functions designed using AJD (top row) and our approach (bottom row). Different from AJD, our energy does not favor matching basis elements which allows for a more flexible design process. See also Fig.~\ref{fig:teaser}.}
  \label{fig:cmp_ajd}
\end{figure}

\subsection{Comparison with AJD~\cite{kovnatsky2013coupled}}
\label{subsec:cmp_to_ajd}

Perhaps closest to our approach is the method that finds approximate joint diagonalized bases of Kovnatsky et al.~\shortcite{kovnatsky2013coupled}. In this work, the authors explore an optimization problem which is conceptually similar to ours. However, there are several key differences between our technique and theirs as we detail below.

In terms of the energy functional, our technique is fundamentally different from theirs. Their approach favors basis elements which diagonalize the LB operator, leading to smooth functions. However, the disadvantage in this point of view is that one implicitly assumes that smooth basis functions span the descriptors subspace. Unfortunately, many practical descriptors that are currently used in functional map pipelines do not fit into this assumption. Indeed, any high frequency signal such as segment information will undergo a low pass filter which may lead to data loss in practice, as we show in Fig.~\ref{fig:feature_err}. In contrast, our method does not favor smooth basis elements and may output high frequency functions, see e.g., Fig.~\ref{fig:faust_pod_vis}. Finally, even when we include Dirichlet terms in our minimization, they are weighted weakly. 

Another significant difference is in the data fidelity term. The formulation in~\cite{kovnatsky2013coupled} and others~\cite{litany2017fully} fixes the associated functional map $C$ to attain a \emph{particular structure}. Namely, they include a term that takes the following form
\begin{align} \label{eq:ajd_fid}
    \tilde{\mca{E}}_{\code{fid}} = \frac{1}{2} |B_1^T F_1 - B_2^T F_2 |_F^2 \ ,
\end{align}
which can be interpreted as setting $C$ to be $C \approx B_2^T B_1$. There are two disadvantages to formulation~\eqref{eq:ajd_fid} which our approach overcomes. First, regularizing the functional map $C$ is not straightforward as in our formulation~\eqref{eq:fmbs_regs}, and may lead to \emph{quartic} expressions in the unknowns $B_j$. Indeed, our formulation allows to independently constrain the bases or the functional map and its inverse, manifesting greater flexibility alongside the natural utilization of state-of-the-art regularizers. Second, our method allows for \emph{general} functional map matrices and thus it increases the search space of solutions when compared with AJD frameworks.

To summarize, our approach generalizes AJD methods in that it combines work on joint diagonalization and functional map optimization in a unified framework. There are three key differences in our technique. First, we consider a much larger search space of solutions as we utilize the Proper Orthogonal Decomposition (POD) modes which are better suited to the given features, and we further allow for general functional map matrices. Second, since we jointly optimize for the functional map and the bases, we can naturally incorporate regularization terms. Finally, on the algorithmic side, AJD approaches facilitate a constrained minimization tool which is inefficient in practice as can be seen in Fig.~\ref{fig:timings_graph} and its convergence is not guaranteed. In contrast, we analyze our approach and show that it is similar to a provably convergent problem.

\begin{figure}[t]
  \centering
  \begin{overpic}[width=\linewidth]{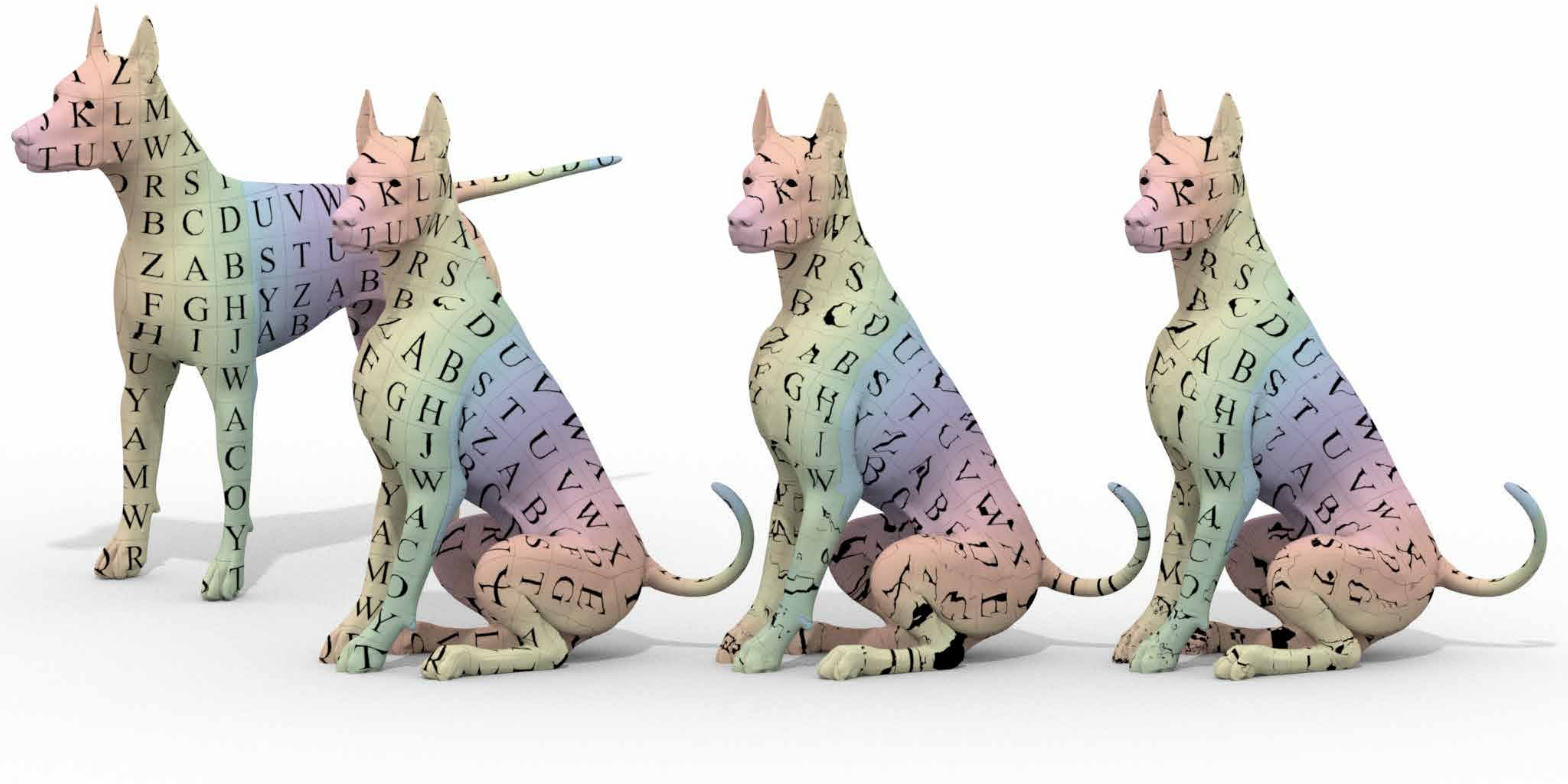}
  \put (23,45) {GT} \put (47.5,45) {AJD} \put (72.5,45) {Ours}
  \end{overpic}
  \caption{We compute functional maps and bases using AJD and our method, and we compare the results to the ground-truth (GT) via texture transfer. }
  \label{fig:cmp_ajd_tex}
\end{figure}

\begin{figure}[b]
  \centering
  \includegraphics[width=\linewidth]{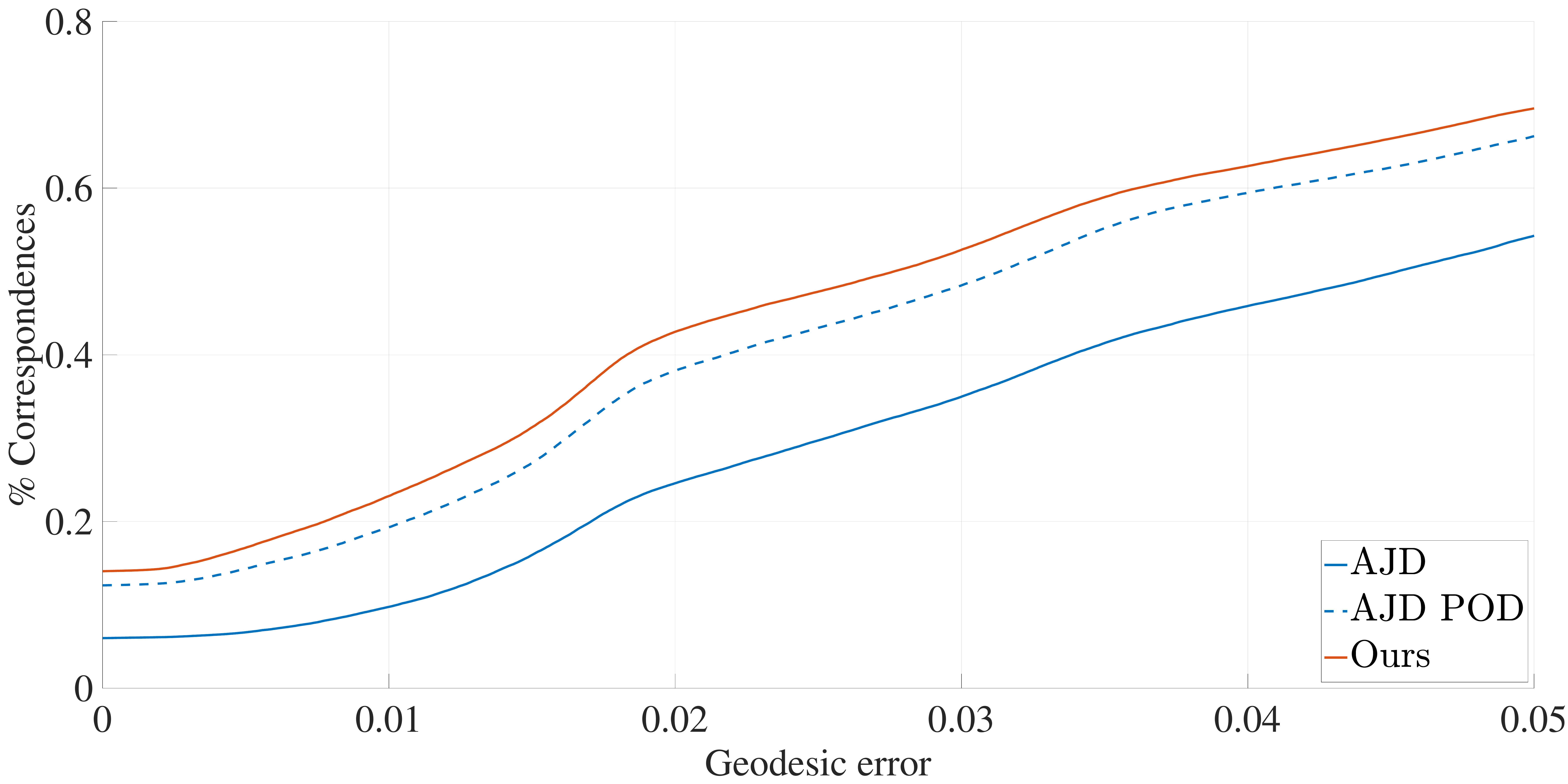}
  \caption{Switching to POD-based design with AJD (blue dash line) yields an improvement over the LB subspaces (blue line). Still, our framework generates correspondences that are more accurate.}
  \label{fig:cmp_ajd_pod}
\end{figure}

\begin{figure*}[t]
  \centering
  \includegraphics[width=\linewidth]{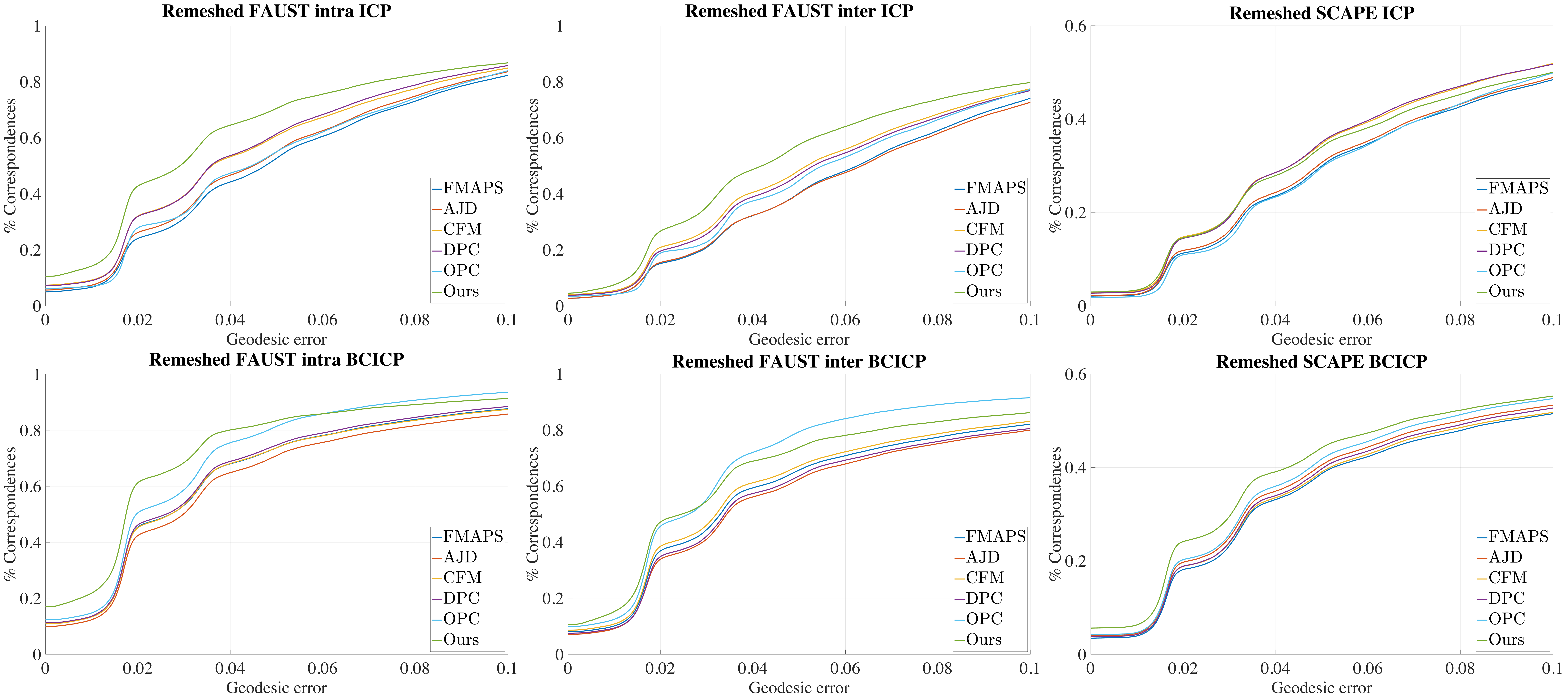}
  \caption{We evaluate our approach in the challenging scenario of shapes with different connectivities and automatically generated descriptors. Our method exhibits very good error measures as compared to state-of-the-art approaches for computing functional maps.}
  \label{fig:seg_wks_cul_err}
\end{figure*}

In addition to this qualitative comparison, we show in Figs.~\ref{fig:cmp_ajd} and \ref{fig:teaser} the differences between the designed basis elements. Indeed, AJD (top row) produces highly consistent basis functions compared to ours (bottom row). However, we believe that this behavior limits the design process significantly, which may lead to less accurate matching results as can be seen in Figs.~\ref{fig:lmk_wks_cull_err} and \ref{fig:seg_wks_cul_err}. Specifically, we select a pair of shapes from TOSCA and visualize the correspondence differences via texture transfer in Fig.~\ref{fig:cmp_map_tex}. Overall, AJD produces reasonable results as compared to the ground-truth (GT). However, various parts of the shape such as head, legs and tail, display large errors. In contrast, our technique was able to accurately match most areas of the shapes including the challenging parts. 

To conclude our qualitative comparison, we modify AJD to use POD modes in their design process instead of the LB eigenfunctions and we plot the cumulative geodesic error that was obtained for FAUST in Fig.~\ref{fig:cmp_ajd_pod}. Indeed, switching to POD modes (blue dashed line) yields a large improvement compared to LB-based AJD (blue line). However, our method (red line) is still significantly more accurate, which can be attributed in part to the state-of-the-art regularization terms we include in our optimization.

\begin{figure}[t]
  \centering
  \includegraphics[width=\linewidth]{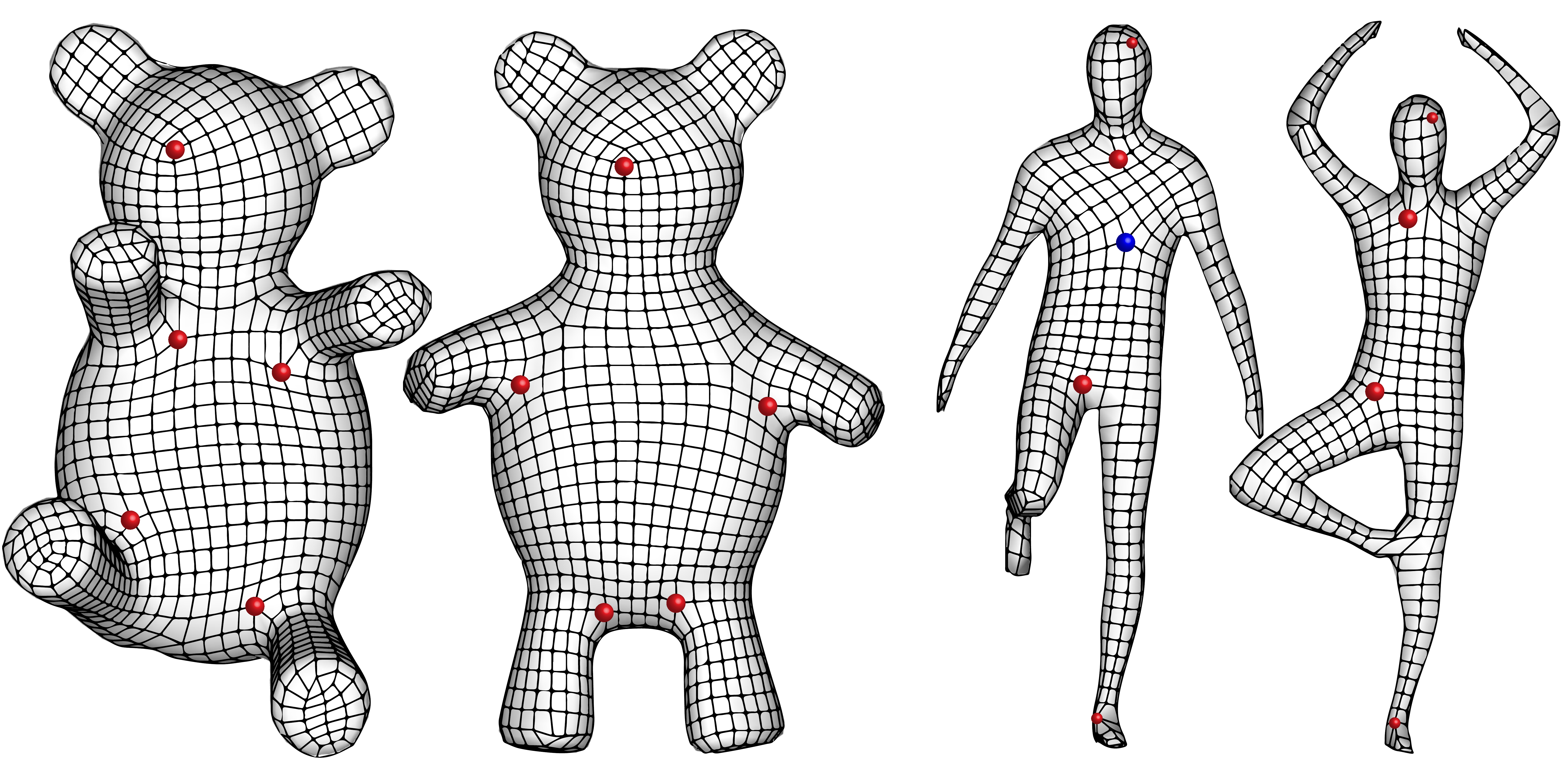}
  \caption{Our technique is particularly suited to methods whose input is a functional map with its associated bases such as~\cite{azencot2017consistent}. We demonstrate the consistent quadrangulations obtained by using their method with input generated by our approach.}
  \label{fig:qremesh}
\end{figure}

\subsection{Consistent quadrangulation and function transfer}

\begin{figure}[t]
  \centering
  \begin{overpic}[width=\linewidth]{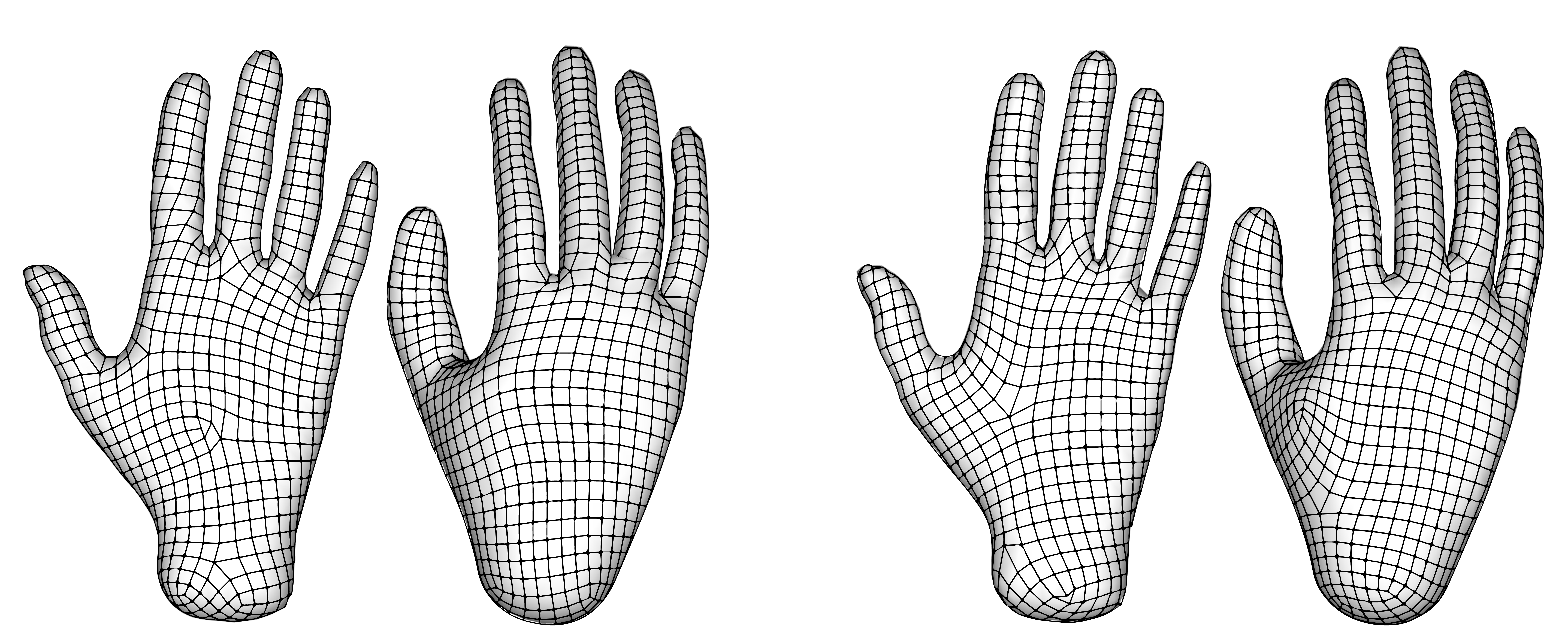}
  \put (19,38) {Fixed LB} \put (68,38) {Designed POD}
  \end{overpic}
  \caption{We compare the quadrangulations produced by using fixed LB bases (left) and designed POD modes (right). Overall, there is a significant improvement in the designed case in terms of isolines alignment, singularity matches and adherence to curvature (see e.g., the index finger, left).}
  \label{fig:cmp_qremesh}
\end{figure}

The increasing interest in the functional map approach over the last few years lead to the development of techniques which can utilize a given functional map directly, without the need to convert it to a point-to-point map. For instance, \cite{azencot2017consistent} proposed an optimization framework for designing consistent cross fields on a pair of shapes for the purpose of generating approximately consistent quadrangular remeshings of the input shapes. Our computed functional map and bases can be directly used within their method to produce quad meshes. In Fig.~\ref{fig:qremesh}, we show an example of the remeshing results of two pairs of shapes having different connectivities (left and right) and genus (right). Still, we obtain highly consistent results as can be seen in the matching singularity points (red spheres). We provide an additional instance of this pipeline in Fig.~\ref{fig:cmp_qremesh} comparing the quadrangulation achieved with fixed LB bases (left) vs. our technique with designed POD modes (right). Indeed, we observe a much better alignment of isolines and singularity points with our approach compared to Fixed LB.

\begin{figure}[b]
  \centering
  \begin{overpic}[width=\linewidth]{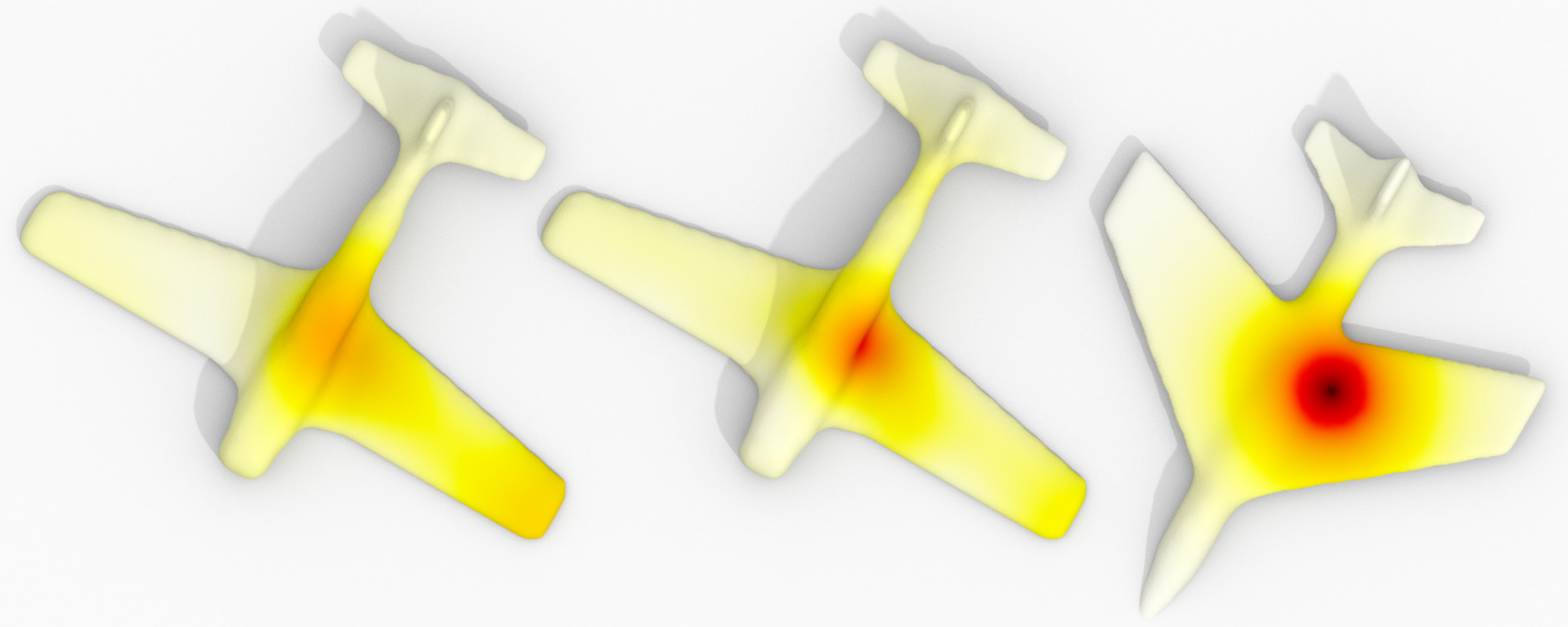}
  \put (5,3) {Standard transfer} \put (40,3) {Product transfer}
  \end{overpic}
  \caption{Mapping Gaussian function between shapes with different connectivities is a challenging task, whose results may be exploited in context of shape matching to construct an accurate correspondence or to improve a given one.}
  \label{fig:func_transfer}
\end{figure}

The last application we consider involves the transfer of scalar valued information between shapes. Recently, \cite{nogneng2018improved} showed that by extending the usual functional basis to include basis products, an improved function transfer can be performed. In Fig.~\ref{fig:func_transfer}, we utilize this pipeline using our functional map and bases to transfer an extremely challenging data given by a localized Gaussian function. Indeed the transfer is improved using the extended basis as the noise is less severe and the maximum is more localized.

\section{Limitations}

One limitation of our framework is related to the dependencies between the given constraints and our choice of dimensionality reducing subspaces $U_j$. Indeed, one can always add the standard LB spectrum to these subspaces. However, we observe that in general, the results may change depending on the particular subspace in use and its size. For instance, while increasing $r$ allows for greater flexibility for representing scalar functions, it also requires more regularization, otherwise unwanted solutions may potentially become local minimizers. Another shortcoming of our approach is that it tends to produce maps that are less smooth compared to those generated with LB bases. This behavior is somewhat expected, as our bases are designed to potentially transfer high frequency information which in turn leads to less uniform correspondences. We leave further investigation of these aspects to future work.

\section{Conclusions and Future Work}

In this paper, we proposed a method for designing basis elements on a pair of triangle meshes along with an associated functional map. Unlike most existing work which utilize the spectrum of the Laplace--Beltrami operator, our technique adopts the Proper Orthogonal Decomposition (POD) modes to reduce the dimensionality of the problem. This choice introduces many degrees of freedom and it significantly extends the space of potential solutions. To effectively solve the problem, we incorporate state-of-the-art regularization terms which promote consistency, isometry and smoothness. Our optimization scheme is based on the Alternating Direction Method of Multipliers (ADMM) and it consists of easy-to-solve linear or Sylvester-type equations. We show that in practice our method behaves well in terms of convergence, and we additionally prove that a similar problem to ours is guaranteed to converge. We evaluate our machinery in the context of shape matching, function transfer and consistent quadrangulation, and we demonstrate that our results yield a significant improvement over state-of-the-art approaches for computing functional maps.

In the future, we would like to characterize the dependencies between the subspaces spanned by the bases to the given constraints and the relation to the functional map. Moreover, we believe that many applications may benefit from the proposed pipeline on a single shape. Namely, generate a self functional map with a set of basis elements defined on the shape. Examples include symmetry detection, fluid simulation and data interpolation, among many other possibilities. On the other hand, extending our framework to handle multiple shapes is an interesting direction as well. Finally, we believe that many of the questions that we consider in our work could benefit from the recent advancements in machine learning with deep neural networks. We plan to investigate how the task of designing a basis and a functional map can be solved using deep learning approaches.



\bibliographystyle{ACM-Reference-Format}
\bibliography{basis-search}

\appendix

\section{Proof of Proposition 1}
\label{app:prop_proof}

We consider a modified version of our problem~\eqref{eq:fmbs_admm} given by
\begin{equation} \label{eq:fmbs_admm2} \begin{aligned}
& \minimize & & \mca{F}(\mca{X},\mca{Z}) \\
& \text{subject to} & & \mca{P}(\mca{X}) + \mca{Q}(\mca{Z}) = 0
\end{aligned} \end{equation}
where $\mca{X} = (B_1,B_1',\tilde{B}_1',B_2,B_2',\tilde{B}_2', C)$ and $\mca{Z} = (Z,B_1'',\tilde{B}_1'',B_2'',\tilde{B}_2'')$ are blocks of variables. Further, the objective function is given by
\begin{align*}
\mca{F}(\mca{X},\mca{Z}) &= \mca{G}(\mca{X}) + \mca{H}(\mca{Z}) \ , \\
\mca{G}(\mca{X}) &= \frac{1}{2} | C \tilde{B}_1'^T F_1 - \tilde{B}_2'^T F_2 |_F^2 \ , \\
\mca{H}(\mca{Z}) &= \frac{\nu}{2} | Z - I |_F^2 \\
&+ \frac{\mu}{2} |B_1''|_{\mca{M}_1}^2 + \frac{\mu}{2} |\tilde{B}_1''|_{\mca{M}_1}^2 \\
&+ \frac{\mu}{2} |B_2''|_{\mca{M}_2}^2 + \frac{\mu}{2} |\tilde{B}_2''|_{\mca{M}_2}^2 \ .
\end{align*}
Finally, the constraints are formed via
\begin{align*}
\mca{P}(\mca{X}) = \begin{pmatrix} B_1^T G_1 \, B_1' \\ B_2^T G_2 \, B_2' \\ B_1-B_1' \\ B_2 - B_2' \\ B_1 - \tilde{B}_1' \\ B_2 - \tilde{B}_2' \end{pmatrix} \ , \quad \mca{Q}(\mca{Z}) = \begin{pmatrix} -Z \\ -Z \\ -B_1'' \\ -B_2'' \\ -\tilde{B}_1'' \\ -\tilde{B}_2'' \end{pmatrix} \ .
\end{align*}


\setcounter{prop}{0}
\begin{prop} 
Under some mild conditions, problem~\eqref{eq:fmbs_admm2} satisfies all the requirements in~\cite{gao2018admm} and thus its ADMM converges.
\end{prop}
 
\paragraph{Proof} 
We need to show that the requirements in Assumption 1 and Assumption 2 in~\cite{gao2018admm} hold. Our variables are updated sequentially in the order $B_1,B_1',\tilde{B}_1',B_2,B_2',\tilde{B}_2', C$ and a single block of $(Z,B_1'',\tilde{B}_1'',B_2'',\tilde{B}_2'')$. We have that $\im(\mca{Q}) \supseteq  \im(\mca{P})$ since the image of $\mca{Q}$ is spanned by the identity matrix in each of the components. The objective function $\mca{F}(\mca{X},\mca{Z})$ is coercive on the feasible set $\{ (\mca{X},\mca{Z}) | \mca{P}(\mca{X}) + \mca{Q}(\mca{Z}) = 0 \}$ since for every variable in $\mca{Z}$ the function behaves as $|x|^2$. This also holds for the variables in $\mca{X}$ because of the constraints. Moreover, the function $\mca{H}(\mca{Z})$ is a strongly convex function because its Hessian is positive definite. Also, every subproblem in the ADMM of~\eqref{eq:fmbs_admm2} is a trivial, linear or Sylvester-type equation and thus it attains its optimal value when $\rho$ is sufficiently large. Finally, our objective term $\mca{G}(\mca{X})$ is differentiable and, in particular, it is lower semi-continuous.

\end{document}